\documentclass[aps,amsmath,amssymb,twocolumn,nofootinbib,prx,superscriptaddress,longbibliography]{revtex4-2}

\usepackage{amsthm}
\usepackage{amsmath}            %serve per le subequazioni
\usepackage{amssymb}            %serve per il simbolo "marchio registrato", \circledR
\usepackage{mathtools}
\usepackage{graphicx}           %serve per le figure eps, ps etcyy
\usepackage[caption=false]{subfig}
\usepackage{multirow}
\usepackage{booktabs}

\usepackage{xcolor}
\usepackage{bm}

\usepackage{soul} % For crossing out words and sentences

\usepackage{tcolorbox}
\usepackage{algpseudocode}
\usepackage{enumitem}

\usepackage[colorlinks,citecolor=blue,urlcolor=blue,bookmarks=false,hypertexnames=true]{hyperref} 
\newcommand{\ket}[1]{| #1 \rangle}
\newcommand{\bra}[1]{\langle #1 |}
\newcommand{\proj}[1]{\ket{#1}\bra{#1}}
\newcommand{\ts}{\otimes}
\newcommand{\id}{\openone}
\newcommand{\tr}{\text{Tr}}

\newcommand{\assemb}[1]{\{#1\}}

\newcommand{\cl}{\text{cl}}
\newcommand{\work}{\overline{W}}
\newcommand{\W}{\mathcal{W}}

\newcommand{\Sup}{\mathcal{S}}

\usepackage[normalem]{ulem}

\newcommand{\comt}[1]{}
\newcommand{\sect}[1]{Sec.~\ref{#1}}
\newcommand{\apx}[1]{Appendix~\ref{#1}}
\newcommand{\eq}[1]{Eq.~(\ref{#1})}
\newcommand{\fig}[1]{Fig.~\ref{#1}}
\newcommand{\tab}[1]{Table~\ref{#1}}

\begin{document}
\title{Maxwell's two-demon engine under pure dephasing noise}

\author{Feng-Jui Chan}
\thanks{Feng-Jui Chan and Yi-Te Huang contributed equally to this work.}
\affiliation{Department of Physics, National Cheng Kung University, 701 Tainan, Taiwan}
\affiliation{Center for Quantum Frontiers of Research \& Technology (QFort), National Cheng Kung University, 701 Tainan, Taiwan}

\author{Yi-Te Huang}
\thanks{Feng-Jui Chan and Yi-Te Huang contributed equally to this work.}
\affiliation{Department of Physics, National Cheng Kung University, 701 Tainan, Taiwan}
\affiliation{Center for Quantum Frontiers of Research \& Technology (QFort), National Cheng Kung University, 701 Tainan, Taiwan}

\author{Jhen-Dong Lin}
% \email{jhendonglin@gmail.com}
\affiliation{Department of Physics, National Cheng Kung University, 701 Tainan, Taiwan}
\affiliation{Center for Quantum Frontiers of Research \& Technology (QFort), National Cheng Kung University, 701 Tainan, Taiwan}

\author{Huan-Yu Ku}
\affiliation{Department of Physics, National Cheng Kung University, 701 Tainan, Taiwan}
\affiliation{Center for Quantum Frontiers of Research \& Technology (QFort), National Cheng Kung University, 701 Tainan, Taiwan}
\affiliation{Faculty  of  Physics,  University  of  Vienna,  Boltzmanngasse 5, 1090 Vienna,  Austria}
\affiliation{Institute  for  Quantum  Optics  and  Quantum  Information  (IQOQI), Austrian  Academy  of  Sciences,  Boltzmanngasse 3, 1090 Vienna,  Austria}

\author{Jui-Sheng Chen}
\affiliation{Department of Physics, National Cheng Kung University, 701 Tainan, Taiwan}
\affiliation{Center for Quantum Frontiers of Research \& Technology (QFort), National Cheng Kung University, 701 Tainan, Taiwan}

\author{Hong-Bin Chen}
\email{hongbinchen@gs.ncku.edu.tw}
\affiliation{Center for Quantum Frontiers of Research \& Technology (QFort), National Cheng Kung University, 701 Tainan, Taiwan}
\affiliation{Department of Engineering Science, National Cheng Kung University, 701 Tainan, Taiwan}

\author{Yueh-Nan Chen}
\email{yuehnan@mail.ncku.edu.tw} 
\affiliation{Department of Physics, National Cheng Kung University, 701 Tainan, Taiwan}
\affiliation{Center for Quantum Frontiers of Research \& Technology (QFort), National Cheng Kung University, 701 Tainan, Taiwan}

\begin{abstract}
The interplay between thermal machines and quantum correlations is of great interest in both quantum thermodynamics and quantum information science. 
Recently, a quantum Szil\'ard engine has been proposed, showing that the quantum steerability between a Maxwell's demon and a work medium can be beneficial to a work extraction task. 
Nevertheless, this type of quantum-fueled machine is usually fragile in the presence of decoherence effects. Therefore, in this work, we tackle this question by introducing a second demon who can access a control system and make the work medium pass through two dephasing channels in a manner of quantum superposition. Furthermore, we provide a quantum circuit to simulate our proposed concept and test it on IBMQ and IonQ quantum computers.%The experimental results are consistent with that obtained from noise simulation, which takes the intrinsic error of the quantum device into account.

\end{abstract}

\maketitle

\section{Introduction}
Due to the development of quantum theory, the microscopic picture of thermodynamics is constructed by reconciling the quantum phenomena and is turned into a new version called quantum thermodynamics. In fact, it is conceived from the necessity to deal with quantum effects, e.g., quantum superposition and quantum correlations, for different scenarios, such as thermal engine~\cite{Scully2011,Rahav2012,Brunner2014,Mitchison2015,HBChen2016,Brandner2017,Manzano2018,Woods2019maximumefficiencyof,Niedenzu2019conceptsofworkin,VanHorne2020,Gluza2021PRXQuantum,Jeongrak2021PRXQuantum,Lu2022PRXQuantum}, protocols for work extractions~\cite{Funo2013,Skrzypczyk2014,Alhambra2016,Elouard2017,Morris2019,Niedenzu2019}, and fluctuations of work~\cite{Allahverdyan2014,Talkner2016,Lostaglio2018}. 
% In particular, recent investigations suggest that quantum resources such as quantum correlations~\cite{Popescu2006,Brando2008,Dillenschneider2009,Perarnau-Llobet2015,Woods2019maximumefficiencyof,Niedenzu2019conceptsofworkin,Chen2019,VanHorne2020,YungerHalpern2022} and quantum coherence~\cite{Scully2003,Lostaglio2015,Cwiklinski2015,Mitchison2015,Korzekwa2016,Santos2019} can be beneficial on thermal machines. %Recently, Beyer \textit{et al.}~\cite{Beyer2019} have shown that quantum steering, which was proposed by Schr\"odinger~\cite{Schrodinger1935} and formalized by Wiseman~\cite{Wiseman2007}, can provide quantum advantage on a quantum Szil\'ard engine, a thermal machine assisted by Maxwell's demon. 
Recently, quantum steering~\cite{Schrodinger1935,Wiseman2007,Cavalcanti2016,Uola2020,Xiang2022}, a type of spatial quantum correlations, has been used to demonstrate quantum advantage on a quantum Szil\'ard engine~\cite{Beyer2019,Ji2022}, a work-extraction machine assisted by Maxwell's demon~\cite{Maxwell1871}.
%In Ref.~\cite{Beyer2019}, the authors
A steering-type inequality is derived in terms of classical limit of the engine's work output, i.e., the maximum extractable work by using classical resources. They further show that some steerable resources can exceed this classical limit, implying that the Maxwell's demon should be certified as a genuinely ‘‘quantum" entity for these cases.
% \cite{Wiseman2007,Cavalcanti2016,Uola2020}
% \orange{HY: I think talking about unsteerable and LHS are too board in here. Maybe classical resource is enough. And the next sentence can be omitted.} %\red{We note that the authors only discussed a particular process of a Szil\'ard engine to test a genuine quantum Maxwell demon.}

% \blue{Recently, coherently controlled quantum channels show that it can not only enhance the communication[blablabla] but also can be used to improve the thermal machine \cite{Simonov2022}blablabla. It consists a control system to manipulate the evolution depending on the state of control system. With the selective measurement on the control system, the evolution shows difference comparing to the classical mixture of quantum channels. We then construct the Szil\'ard engine combine with the idea of quantum steering and superposition of quantum channels~\cite{Oi2003,Gisin2005,Chiribella2019quantum,Loizeau2020,Abbott2020,Kristjansson2020resource,Rubino2021,Henderson2020,Ban2020relaxation,MiguelRamiro2021,Siltanen2021}, one kind of coherently controlled quantum channels. To achieve the superposition of the quantum channels, the engine needs another demon to prepare the control system and perform the selective measurement. We show that with the second demon's assistance, there will be an enhancement on the average extracted work. 
% }
In contrast to the quantum correlations for spatially separated systems, Leggett and Garg proposed the concept called (non-)Macrorealism and the well-known Leggett-Garg inequality~\cite{Leggett85,Joarder2022PRXQuantum}, suggesting that the non-classical features can also be revealed in temporal evolutions of quantum systems (see Ref.~\cite{Emary14} for a comprehensive review). The nonclassical properties in time domain are now termed ‘‘temporal quantum correlations", which are beneficial to several quantum information tasks, such as certifying quantum memory~\cite{Ku2021,Vieira2022temporal} and self-testing quantum measurement~\cite{Maity2021}.

As aforementioned, Ref.~\cite{Beyer2019} investigated the influence of spatial quantum steerability on a heat engine. Here, we propose a  mirror of this steering heat engine by considering temporal quantum steering~\cite{Chen2014,Li2015,Chen2016,Ku2018,Lin2021}. More specifically, in each round of the experiment, the Maxwell's demon (say Alice for convenience) performs a measurement on the work medium taken from a heat bath. She then sends the work medium to Bob through a quantum channel, so that Bob can charge his battery by extracting energy from the work medium. %\red{We note that this transformation process can be implemented by i.e., communication line~\cite{Cirac1997}.}
We also derive a temporal steering inequality in terms of the classical limit of the extractable work, which can be numerically computed via semidefinite program (SDP). Therefore, we can identify useful temporal steerable resources that can demonstrate quantum advantage for the work extraction task.

In practice, the quantum channel between Alice and Bob could suffer from unwanted noise that further degrades the temporal steerability as well as the engine's performance. To tackle this problem, we adopt the recent developed scheme called superposition of quantum channels, which has been utilized in quantum information science~\cite{Guerin2016,Chiribella2019quantum,Abbott2020,kristjansson2020resource,Ban2021two,Rubino2021,Lin2022,Lee2022}.
%To enhance the extractable work when the dephasing channel is present, which has been used in quantum information~\cite{Guerin2016,Abbott2020,Henderson2020,kristjansson2020resource,Foo2021,Ban2021two,Rubino2021}. 
This approach involves two or several channels placed in parallel and introduces a control system to decide which channel for the work medium to pass through.
We consider that the control system can be accessed by another demon, Charlie. He can prepare the control system in a quantum superposition state so that the work medium can also pass through the channels in a manner of quantum superposition. Before Bob receives the work medium, Charlie can perform a selective measurement on the control system and discard the unwanted results. 

In this work, we consider a superposition of two pure dephasing channels. With Charlie's assistance, one can observe an enhancement of the extractable work in comparison with the case that involves only one dephasing channel. Moreover, we implement the two-demon engine on IBMQ and IonQ quantum computers~\cite{IBMWeb,Berke2022,Strikis2021PRXQuantum,Pogorelov2021PRXQuantum}. We find that, even in the presence of the intrinsic errors of the quantum computers, the results still demonstrate significant enhancement with the help of the superposition of quantum channels and the second demon. To further analyze the effect of the intrinsic errors, we also consider a Markovian noise model to numerically simulate devices' imperfections. We find that the results for IBMQ are in good agreement with our noise simulations; while those for IonQ contain unexpected oscillatory behavior, suggesting a non-Markovian nature of the device.

\section{Single-Demon Heat Engine}\label{sec:temporal_steering_heat_engine}
\begin{figure}[!htbp]
    \centering
    \includegraphics[width=1\linewidth]{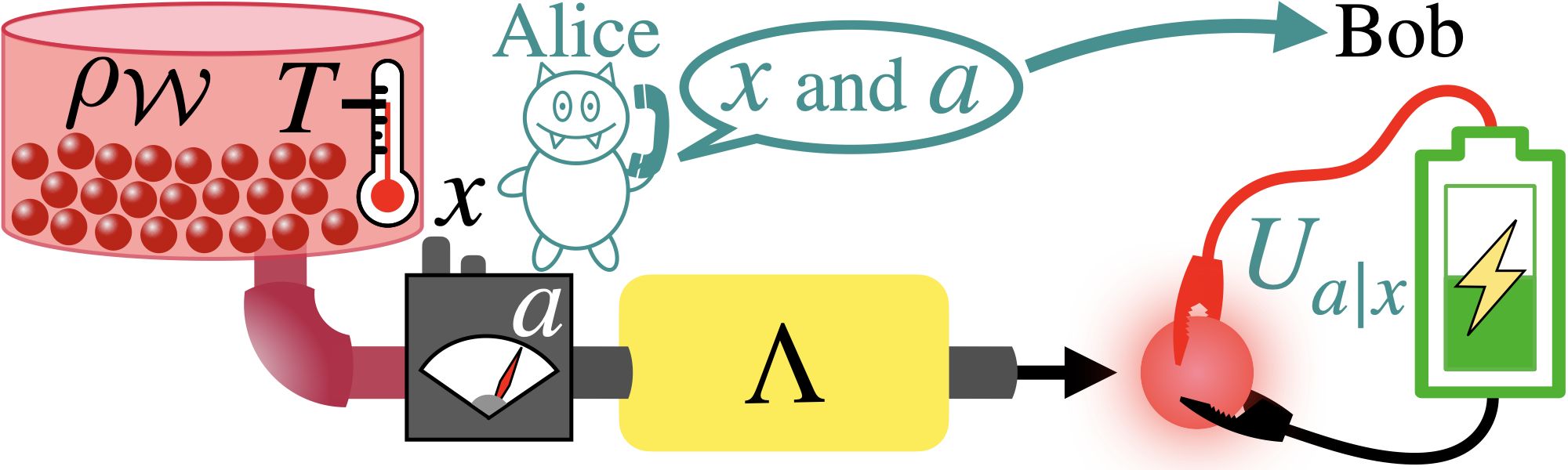}
    \caption{Schematic illustration of the single-demon heat engine. Alice (a Maxwell's demon) takes a work medium $\W$ from a thermal bath at temperature $T$. She then performs a measurement labeled by $x$ and obtains the corresponding outcome $a$. Then, Alice informs Bob of ($a$,$x$) through a classical communication and sends the post-measurement work medium through a pure dephasing channel $\Lambda$ to Bob. After that, Bob chooses a suitable work extraction operation $U_{a|x}$ based on Alice's information. 
    }
    \label{fig:scenario1}
\end{figure}
In this section, we provide a detailed description of the work-extraction engine with only one Maxwell's demon (Alice). Also, we derive the temporal steering inequality in terms of the classical limit of the engine's work output.

Let us start from the process to extract work from a work medium. We consider that the work medium $\W$ is modeled as a two-level system with the Hamiltonian $\widehat{H}_\W=\hbar \omega \proj{1}$, where $\hbar\omega$ is the energy difference between the excited state $\ket{1}$ and the ground state $\ket{0}$.
%we consider two-level systems with a Hamiltonian $\widehat{H}_\W=\hbar \omega \proj{1}$ as the work medium $\W$. Here, $\hbar \omega$ is the energy difference between the excited state $\ket{1}$ and the ground state $\ket{0}$ of $\widehat{H}_\W$. 
In general, a work extraction operation is a process that transfers the energy from a work medium to a battery. As reported in Refs.~\cite{Skrzypczyk2014,Alhambra2016}, an ideal work extraction process can be described by a local unitary evolution acting on the work medium $\W$, because it does not decrease the entropy of $\W$. If the quantum state of the work medium is available, one can then design a suitable work extraction operation for this particular state. When the work medium is prepared in a pure state $\ket{\psi}$, then the optimal work extraction strategy is described by a local unitary $U$ such that $U \ket{\psi}=\ket{0}$. This idea can be generalized to the situation where the work medium is prepared in a mixed state. A mixed state can be interpreted as a statistical mixture of pure states. In fact, for a given mixed state $\rho_\W$, there are infinite possibilities of pure state decomposition $D=\lbrace p_k, \proj{\psi_k} \rbrace$ such that $\sum_k p_k \proj{\psi_k}=\rho_\W$. If a Maxwell's demon can reveal the knowledge of this decomposition, one can then design a set of optimal work extraction operations $\lbrace U_k \rbrace$. Therefore, the average extracted work can be expressed as $\work=\sum_k p_k \Delta W_k$, where 
\begin{equation}
    \Delta W_k=\tr\left[\widehat{H}_\W\proj{\psi_k}\right]-\tr\left[\widehat{H}_\W U_k \proj{\psi_k} U_k^\dagger \right].
\end{equation}

Now we introduce the single-demon heat engine, as shown in \fig{fig:scenario1}. A thermal bath can supply unlimited copies of work medium $\W$, which is initialized in the Gibbs state at temperature $T$. Alice (the demon) performs a measurement (labeled by $x$) on $\W$ with the associated outcome (labeled by $a$). The work medium with the post-measurement state $\rho_{a|x}$ is then sent to Bob via a noisy quantum channel $\Lambda$. In addition, Alice informs Bob of $(a,x)$ through a classical communication so that Bob can apply a unitary operator $U_{a|x}$ to $\W$ to extract work according to Alice's message. 
%$U_{a|x}$ can map the post-measurement states onto the $\ket{0}$ to extract work from work medium $\W$. 
Thus, the extracted work for the measurement $x$ with the outcome $a$ is given by
\begin{equation}\label{eq:delta_w_a_x}
    \Delta W_{a|x}=\tr\left[\widehat{H}_\W \Lambda(\rho_{a|x})\right]-\tr\left[\widehat{H}_\W U_{a|x} \Lambda(\rho_{a|x}) U_{a|x}^{\dagger}\right].
\end{equation}
%\blue{Since it requires at least two measurements do not commute with each other to demonstrate quantum property,} 
The average extracted work can then be expressed as
\begin{alignat}{1}\label{eq:avg_extracted_work}
    \work_\Lambda=&\sum_{a,x} p(x)p(a|x) \Delta W_{a|x}\notag\\
    =& \sum_{a,x} \tr\left[p(a|x) F_{a|x} \Lambda(\rho_{a|x})\right],
\end{alignat}
% \orange{
% Here, $p(x)=1/2$ is the probability for the choice of the measurement $x$.
% Moreover, we set  $F_{a|x}:=p(x)(\widehat{H}_\W-U_{a|x}^{\dagger}\widehat{H}_\W U_{a|x})$ as shorthands of the equation.
% % \orange{HY: no physical explaintions of $\sigma$} 
% %In this case, to simply the folowing discussion, we suppose that $\W$ is thermalized to infinite temperature, i.e., $\rho_\W=\id/2.$ 
% In \tab{tab:operators}, we summarize the probability distribution $p(a|x)$ post-measurement state $\lbrace \rho_{a|x} \rbrace$, the operator $\lbrace F_{a|x} \rbrace$.}
where the shorthand $F_{a|x}:=p(x)(\widehat{H}_{\W}-U_{a|x}^{\dagger} \widehat{H}_{\W} U_{a|x})$ is adopted for convenience, $p(x)$ is the probability of Alice's choice on measurements, and $p(a|x)$ is the probability of obtaining the outcome $a$ conditioned on the measurement $x$. The work-extraction unitaries $\lbrace U_{a|x} \rbrace_{a,x}$ can be optimized for the post-measurement states $\lbrace \rho_{a|x} \rbrace_{a,x}$, i.e., $U_{a|x} \rho_{a|x} U_{a|x}^\dagger=\proj{0}$ for all $a$ and $x$. However, the presence of the noise in the quantum channel $\Lambda$ invalidates the optimal extraction process;
consequently, the average extracted work will be impaired due to the noise. Note that the average extracted work $\work_\Lambda$ can be regarded as a linear function of the so-called temporal steering assemblage $\{p(a|x)\Lambda(\rho_{a|x})\}_{a,x}$, which is a widely accepted terminology for characterizing one-sided device-independent nature of steering scenarios~\cite{Branciard2012,NC2021:BYadin,Zhao2020,Slussarenko2022}. In other words, one can relax the assumptions of Alice's measurement devices.

Let us now present a concrete example. 
%\red{to show how a quantum Szilard engine works}. %which will be useful when we introduce the second-demon scenario.
In each round of the work extraction task, Alice receives a work medium in the Gibbs state at the infinite temperature, i.e., $\rho_\W=\id/2$, from the bath and randomly performs one of the two measurements $\sigma_z=\proj{0}-\proj{1}$ or $\sigma_x=\ket{0}\bra{1}+\ket{1}\bra{0}$ on $\W$ with equal probability. She then sends the $\W$ with the post-measurement state $\rho_{a|x}$ to Bob through a pure dephasing channel $\Lambda$ described by 
\begin{equation}\label{eq:dephasing_channel}
    \Lambda(\rho)= \left( 1-\frac{\gamma}{2} \right)\rho + \frac{\gamma}{2} \sigma_z \rho \sigma_z ,
\end{equation}
where $\gamma$ denotes the dephasing strength. In addition, she informs him of the message $(a,x)$ via a classical communication. 
After receiving the message, Bob then applies $U_{a|x}$ to the work medium $\W$. The corresponding $U_{a|x}$, $F_{a|x}$, and the measurement settings are summarized in \tab{tab:operators}. For this setting, the average extracted work is given by 
\begin{equation}
    \work_\Lambda=\frac{2-\gamma}{4}~\hbar \omega.
\end{equation}
As expected, the average extracted work decreases monotonically when the dephasing strength $\gamma$ increases.
We note that, since $\Lambda$ is Gibbs preserving~\cite{Janzing2000,Faist2015_1,Faist2015_2,Lostaglio2019,Hsieh2021}, it does not change the thermal state of $\W$ on average. Thus, the work medium remains the same Gibbs state on average after the delivery, i.e., $\sum_a p(a|x)\Lambda(\rho_{a|x})=\rho_\W$ for all $x$.

\begin{table}[!htbp]
  \caption{
      A list of Alice's measurement results, $p(a|x)$ and $\rho_{a|x}$, and the work extraction procedure (described by $U_{a|x}$ and $F_{a|x}$) on Bob's side. Here, $\ket{\pm}=(\ket{0}\pm\ket{1})/\sqrt{2}$, $\sigma_z=\proj{0}-\proj{1}$, $\sigma_x=\ket{0}\bra{1}+\ket{1}\bra{0}$, and $H=(\sigma_z + \sigma_x)/\sqrt{2}$ denotes the Hadarmard transform. 
  }
  \label{tab:operators}
  \begin{tabular}{ccccc}
      \hline\hline
      
      $x$ & \multicolumn{2}{c}{$\sigma_z$} & \multicolumn{2}{c}{ $\sigma_x$}\\[1pt]
      
      $p(x)$ & \multicolumn{2}{c}{$0.5$} & \multicolumn{2}{c}{$0.5$}\\[1pt]
      \cmidrule(l){2-3} \cmidrule(l){4-5}
      
      $a$ & $+1$ & $-1$ & $+1$ & $-1$\\[1pt]
      
      \hline
      
      $p(a|x)$ & $0.5$ & $0.5$ & $0.5$ & $0.5$\\[3pt]
      
      $U_{a|x}$ & $\id$ & $\sigma_x$ & $H$ & $\sigma_x H$ \\[3pt]
      
      $\rho_{a|x}$ & $\proj{0}$ & $\proj{1}$ & $\proj{+}$ & $\proj{-}$ \\[3pt]
      
      $F_{a|x}$ & $~0~$ & $~-\frac{\hbar \omega}{2}\sigma_z~$  & $~-\frac{\hbar \omega}{4}(\sigma_z - \sigma_x)~$ & $~-\frac{\hbar \omega}{4}(\sigma_z + \sigma_x)~$ \\[3pt]
      
      \hline\hline
  \end{tabular}
\end{table}

Let us turn to the description of classical strategy. In classical world, one can perform non-invasive measurement to reveal physical properties of a system without disturbing its state and its consequent dynamics~\cite{Leggett85}. Suppose that there exists a hidden variable $\lambda$ associated with a classical randomness $p(\lambda)$, and for each moment the work medium $\W$ is described by a predetermined hidden state $\rho_\lambda$. If Alice performs non-invasive measurements, where she only reveals the hidden variable $\lambda$ without changing the state of $\W$, her measurements can be described by a classical post-processing $p(a|x,\lambda)$. In this case, the temporal steering assemblage received by Bob can be described by a hidden state (HS) model, i.e.,
\begin{equation}\label{eq:HS}
    \sigma_{a|x}^{\text{HS}}=\sum_\lambda p(\lambda)p(a|x,\lambda)\rho_\lambda~~\forall~a,~x.
\end{equation}
Therefore, given a work-extraction protocol, which is described by $\lbrace F_{a|x} \rbrace_{a|x}$ and the Gibbs state $\rho_\W$, the maximal value of the average extracted work attainable by classical assemblages is given by
%Now, given a set of operators $\{F_{a|x}\}_{a,x}$ and the Gibbs state $\rho_\W$, one can define the classical limit of the average extracted work as
\begin{align}\label{eq:Wcl}
\work_\cl&=\max_{\assemb{\tilde{\sigma}_{a|x}}\in \mathcal{C}}~\sum_{a,x}\tr\left[F_{a|x} \tilde{\sigma}_{a|x}\right],\nonumber\\
&\text{s.t.}~\sum_a\tilde{\sigma}_{a|x}=\rho_\W~\forall~x,
\end{align}
where $\mathcal{C}$ denotes the set of all HS models. The constraint in the second line of \eq{eq:Wcl} is introduced because the HS model should also preserve the same Gibbs state on average. Moreover, \eq{eq:Wcl} can be considered as a convex optimization problem, because $\mathcal{C}$ is a convex set. In Appendix~\ref{apx:classical_bound}, we show that one can recast the classical limit of average extracted work in \eq{eq:Wcl} as a semidefinite program (SDP) , which can be solved numerically~\cite{PICOS}. Therefore, when a temporal steerable resource can exceed this classical limit, i.e., $\overline{W}_\Lambda \geq \overline{W}_\mathrm{cl}$, Bob can be convinced that Alice is a genuinely ‘‘quantum" Maxwell's demon.

For the aforementioned concrete example, as summarized in \tab{tab:operators}, the classical limit of the $\work$ is
\begin{equation}\label{eq:Wcl_value}
    \work_\cl\approx0.354~\hbar \omega
\end{equation}
(see also Appendix~\ref{apx:classical_bound} for the optimal solution of HS model in this case). Note that, the above classical limit can be violated if $\gamma < \gamma_\text{TH} \approx 0.586$, where $\gamma_{\text{TH}}$ is the quantum-to-classical transition threshold for the dephasing channel in \eq{eq:dephasing_channel}.
%Because $\mathcal{L}$ is a convex set, Eq.~(\ref{eq:Wcl}) is in general a convex optimization problem and can be solved by the SDP presented in Appendix~\ref{apx:classical_bound}. In the case, as shown in \tab{tab:operators}, the classical limit of the $\work$ is
%\begin{equation}\label{eq:Wcl_value}
%    \work_\cl=\frac{1}{2\sqrt{2}}~(\hbar \omega),
%\end{equation}
%Note that, the classical limit in \eq{eq:Wcl_value} can be violated if $\gamma < \gamma_{\Lambda} = 2-\sqrt{2}$, where $\gamma_{\Lambda}$ is the quantum-to-classical transition threshold for $\Lambda$.

It is worthwhile to note that the SDP method presented in Appendix~\ref{apx:classical_bound} can obtain the same values of the classical limit in the cases proposed in Refs.~\cite{Beyer2019,Ji2022}.
Nevertheless, their derivations only work for restricted considerations of $p(x)$ as well as the numbers of measurements $x$ and outcomes $a$.
Here, we generalize their considerations by providing an efficient approach to compute the value of classical limit in arbitrary work extraction tasks. 

%\red{HY:[The following can be omit. And I do not know whether thermodynamical steering witness is a good name. Maybe engine steering witness?]}
Further, in comparison with the standard steering inequality (see Ref.~\cite{Uola2020} for instance), the classical limit $\work_\text{cl}$ in \eq{eq:Wcl} endows the mathematical hyperplane theory with a physical interpretation in a thermodynamical fashion. Thus, $\work_\text{cl}$ can be seen as a thermodynamical temporal steering witness.
%provides a hyperplain theory with the thermodynamical fashion. 
However, the thermodynamical steering witness can not certify all steerable resource %$\{\sigma_{a|x}\}_{a|x}$ 
because the constitution of the Hamiltonian and the unitary cannot in general represent all positive semidefinite operator, which is used to construct a steering witness.

\section{Two-Demon heat engine}\label{sec:superposition_of_quantum_channels}
\begin{figure}[!htbp]
    \centering
    \includegraphics[width=1\linewidth]{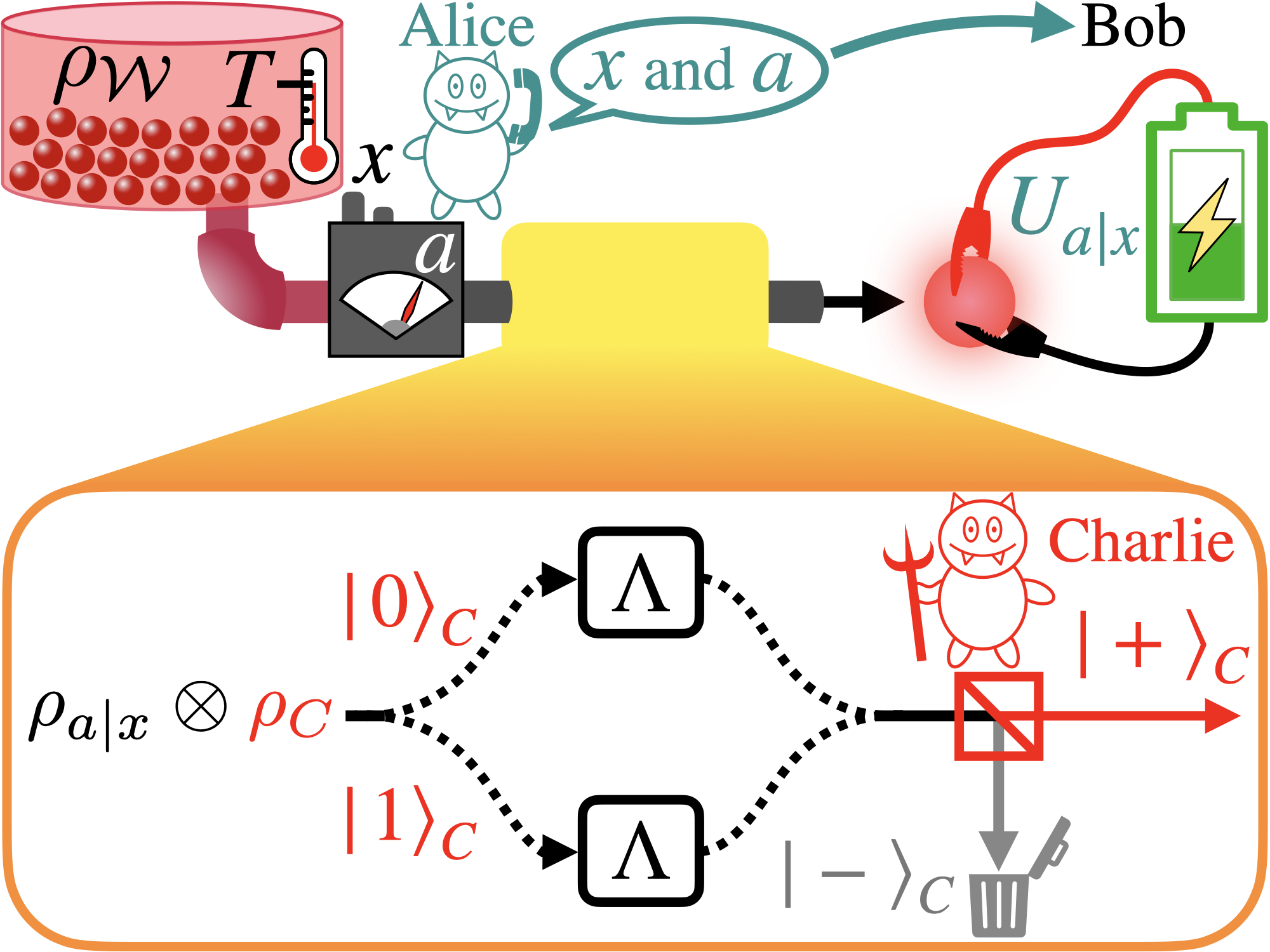}
    \caption{Schematic illustration of the two-demon heat engine. The second demon, Charlie, can prepare a state of control system $\rho_C$ to determine which of the two channels the work medium to pass through and perform a selective measurement on the control system. If the control system is prepared in a superposition state, the work medium can pass through two dephasing channels in a manner of superposition.}
    \label{fig:scenario2}
\end{figure}
%\blue{In this section, we will first introduce the second demon and described the engine with two demons and the superposition of quantum channels, as shown in \fig{fig:scenario2}.

%Before we introduce the heat engine with two demons, we first recall the Maxwell's demon in classical thermodynamics. It is described as an entity can determine the system dynamics by using the information obtained from measurement. Based on this concept, we introduced another demon named Charlie. He can prepare the control qubit in superposition state, and only allows the work medium to reach Bob when the work medium is measured in a specific state.

% As shown in \fig{fig:scenario2}, we now consider the quantum-controlled superposition of dephasing channels, and introduce another Maxwell's demon, Charlie. 
We have shown the detrimental effects of the noise in the quantum channel $\Lambda$ on the work extraction task. To quench the noise in $\Lambda$ requires substantial efforts, which are typically formidable obstacles in various branches of quantum technology. Instead of tackling the noise by engineering the quantum channel $\Lambda$, we circumvent this obstacle by the approach of the superposition of quantum channels, as shown in \fig{fig:scenario2}.

More specifically, instead of a single quantum channel $\Lambda$, we consider a superposition of two channels, which involves an additional control system $C$ that can be accessed by the second demon, Charlie.
%two pure dephasing channels and introduce a control system $C$ which can be accessed by Charlie. 
The state of $C$ determines which channel for $\W$ to pass through. Charlie also performs a selective measurement on $C$ to condition the dynamics of the work medium before reaching Bob.
%a selective measurement on $C$ before Bob receives the work medium. 
In the following, we will show that if the control system is prepared in a superposition state, one can observe a clear enhancement of the average extracted work compared with the case of the single-use of the dephasing channel discussed in the previous section. %In this case, Charlie can be regarded as the second Maxwell's demon in the quantum Szil\'ard engine. 

Before elaborating the superposition of quantum channels~\cite{Abbott2020}, it will be helpful to specify the implementation of a single quantum channel. 
According to Stinespring dilation theorem~\cite{Stinespring1955,Kraus1983,Wilde2017}, any channel with Kraus operators $\{K_k\}_k$ can be implemented by a unitary $V_{\W,E}$ acting on the work medium $\W$ and an environment $E$.
% \begin{equation}\label{eq:stinespring_dilation}
%     V_{\W,E}(\ket{\psi}_\W\ts\ket{\varepsilon}_{E})=\sum_k K_k\ket{\psi}_\W\ts\ket{k}_{E},
% \end{equation}
% where $\{\ket{k}_{E}\}_k$ is a set of orthonormal states of the environment $E$, and $\ket{\psi}$ ($\ket{\varepsilon}$) is the initial state of $\W$ ($E$). 
For the dephasing channel \eq{eq:dephasing_channel}, it can be implemented by the following $V_{\W,E}$:
\begin{equation}\label{eq:dephasing_model}
    \begin{aligned}
    V_{\W,E}\ket{0}_\W\ket{0}_{E} &= \sqrt{1-\frac{\gamma}{2}}\ket{0}_\W\ket{0}_{E}-i\sqrt{\frac{\gamma}{2}}\ket{0}_\W\ket{1}_{E},\\
    V_{\W,E}\ket{1}_\W\ket{0}_{E} &= \sqrt{1-\frac{\gamma}{2}}\ket{1}_\W\ket{0}_{E}+i\sqrt{\frac{\gamma}{2}}\ket{1}_\W\ket{1}_{E}.
    \end{aligned}
\end{equation}
If $E$ is initialized in the state $\ket{0}_E$, the corresponding Kraus operators are $K_0=\sqrt{1-\gamma/2}~\id$ and $K_1=-i\sqrt{\gamma/2}~\sigma_z$.

To implement the superposition of two quantum channels, we introduce two independent environments, $E_0$ and $E_1$, giving rise to the two channels to be superposed.
Along with a control qubit $C$, Charlie can determine which environment to interact with according to the state of $C$ by the global unitary
\begin{equation}\label{eq:total_trans}
    V_{\text{T}}=\proj{0}_C\ts V_{\W,E_0}+\proj{1}_C\ts V_{\W,E_1}.
\end{equation}
%Now, as shown in Fig.~\ref{fig:scenario2}, we consider two pure dephasing channels and introduce a two-dimensional system $C$. \blue{If $C$ is in the state $\ket{j}_C$, $\W$ goes through the pure dephasing channel, i.e., $\W$ interacts with the environment $E_j$.} In other words, $C$ can be regarded as a quantum control and determines which channel for $\W$ to pass through. 
Particularly, if $C$ is prepared in a superposition state $\ket{+}_C=(\ket{0}_C+\ket{1}_C) /\sqrt{2}$,
$\W$ can pass through the two channels in a manner of quantum superposition. 
% More formally, the total unitary transformation for $C$, $\W$, and the environments ($E_0$ and $E_1$) is written as
% \begin{equation}\label{eq:total_trans}
%     V_{\text{t}}=\proj{0}_C\ts V_{\W,E_0}+\proj{1}_C\ts V_{\W,E_1}.
% \end{equation}
%By tracing out the environments, the reduced state of $C$ and $\W$ then reads
Then the joint state of $C$ and $\W$ evolves according to
\begin{equation}\label{eq:control-medium-output-state}
\begin{aligned}
    \rho_{C\W}=&\tr_{E_0,E_1}\left[V_{\text{T}}\left(\proj{+}_C\ts\rho\ts\proj{0}_{E_0}\ts\proj{0}_{E_1}\right)V^\dagger_{\text{T}}\right]\\
    =&\frac{1}{2}\left[\id\ts\Lambda(\rho)+\sigma_x \ts K_0\rho K_0^\dagger\right].
\end{aligned}
\end{equation}
%where, $\rho$ is an arbitrary state of $\W$.
%Let us now consider the superposition scenario described in Fig.~\ref{fig:scenario2}, where Charlie performs a selective measurement on $C$ before Bob receives the work medium $\W$. In our work, we consider 

Before sending $\W$ to Bob, Charlie will perform a selective measurement on $C$. For a concrete example, we assume that Charlie measures the observable $\sigma_x$ on $C$ and only selects the outcome associated with the eigenstate $\ket{+}_C$; while the one associated with the eigenstate $\ket{-}_C$ is discarded.
%and discards the outcome associated with the eigenstate $\ket{-}_C$. 
The normalized post-measurement state of $\W$ is given by %\red{HY: If you only follow this formula, the no-signaling in time cannot be hold. It is because Eq. 13 is non linear. If you want to keep no-signaling in time, you must follow the steps in Refs.~\cite{Rodrigo2015,Nery2020,Gupta2021,Ku2022}. Note, we discuss this before. }
\begin{equation}\label{eq:superposed_channel}
    \Sup(\rho)=\frac{\Lambda_+(\rho)}{\tr\left[\Lambda_+(\rho)\right]},
\end{equation}
where
\begin{equation}\label{eq:Lambda_plus}
\begin{aligned}
    \Lambda_+(\rho)&=\tr_C\left[(\proj{+}\ts\id)\rho_{C\W}(\proj{+}\ts\id)^\dagger\right]\\
    &=\frac{1}{2}\left[\Lambda(\rho)+K_0\rho K_0^\dagger \right].
\end{aligned}
\end{equation}
Note that \eq{eq:superposed_channel} is in general a non-linear equation. Nevertheless, $\Sup(\rho)$ is still a linear process in this case because $\tr\left[\Lambda_+(\rho)\right]=1-\gamma/4$ for arbitrary input state $\rho$. We refer the reader to see Refs.~\cite{Rodrigo2015,Nery2020,Gupta2021,Ku2022} when the most general cases are considered. Therefore, the effective evolution of $\W$ under the superposition scenario can be written as 
\begin{equation}
    \Sup(\rho)= \left( 1-\frac{\gamma'}{2} \right)\rho + \frac{\gamma'}{2} \sigma_z \rho \sigma_z,
\end{equation}
where $\gamma'=2\gamma/(4-\gamma)$. It is crucial to note that the overall effect of the superposition of two pure dephasing channels is equivalent to a single pure dephasing channel with dephasing strength $\gamma' \leq \gamma$, for $0\leq \gamma \leq 1$. Therefore the effect of the noise in a single channel is quenched.

%One can observe that, in this case, the superposition of two pure dephasing channel can also be considered as a pure dephasing channel in \eq{eq:dephasing_channel} with different dephasing strength $\gamma'$.

\begin{figure}[!htbp]
    \centering
    \includegraphics[width=0.98\linewidth]{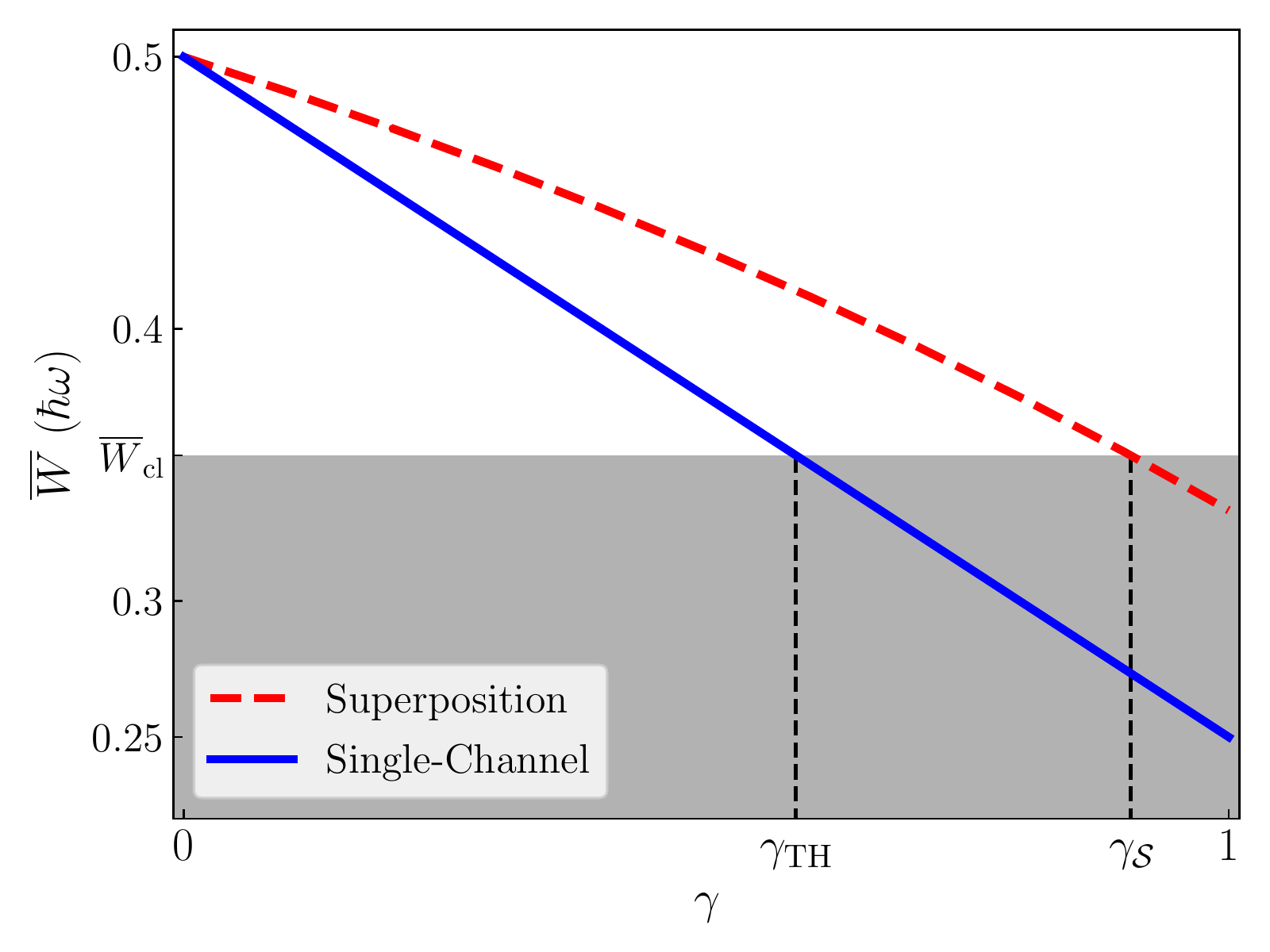}
    \caption{
        The average extracted works for pure dephasing channels. The blue solid line and red dashed line represent the $\work$ obtained from single-channel and superposition scenario, respectively. One can observe that the average extracted work can be enhanced when the second demon (Charlie) superposes two dephasing channels, i.e., $\work_{\Sup}\geq \work_{\Lambda}$. Here, $\gamma_\text{TH} \approx 0.586$ and $\gamma_{\Sup} \approx 0.906$ are the quantum-to-classical transition thresholds.
    }
    \label{fig:ideal_case}
\end{figure}

% \begin{equation}
% \begin{aligned}
%     \Sup(\rho)&=\frac{\left(1-\frac{\gamma}{2}\right)\id\rho\id+\frac{\gamma}{4}\sigma_z\rho\sigma_z}{1-\frac{\gamma}{4}}\\
%     &=\sum_k K_k'\rho K'_k^\dagger.
% \end{aligned}
% \end{equation}
% Here, the Kraus operators are  $K'_0=\sqrt{1-\gamma'/2}~\id$ and $K'_1=\sqrt{\gamma'/2}~\sigma_z$ with $\gamma'=2\gamma/(4-\gamma)$. One can observe that $\Sup$ is an effective pure dephasing channel with dephasing strength $\gamma'$; and hence, for this particular implementation, $\Sup$ also meets the requirement of Gibbs preservation.
% Note that $\Lambda_+$ is in general a complete-positive and trace-non-increasing map.
% \blue{Followed by the resource theory of the steering},
% the temporal steering assemblage then becomes~\cite{Rodrigo2015,Nery2020,Gupta2021,Ku2022}
% \begin{equation}\label{eq:superposed_channel_assemb}
%     \Sup(\sigma_{a|x})=\frac{\Lambda_+(\sigma_{a|x})}{\tr\left[\Lambda_+(\rho_{\W})\right]}~~\forall~~a,x.
% \end{equation}
Finally, we apply this approach to the work extraction task, and the average extracted work in this case reads 
% \begin{equation}
%     \work_\Sup=\sum_{a,x} \tr\left[ F_{a|x} \Sup(\sigma_{a|x})\right].
% \end{equation}

\begin{equation}\label{eq:avg_extracted_work_superposed}
\begin{aligned}
    \work_{\Sup}&=\sum_{a,x} \tr\left[ p(a|x) F_{a|x} \Sup(\rho_{a|x})\right]\\
    &=\frac{2-\gamma'}{4}~\hbar\omega.
\end{aligned}
\end{equation}
The transition threshold in this case is given by $\gamma_\Sup \approx 0.906$. The comparison of the average extracted works between single-channel and superposition scenario is presented in \fig{fig:ideal_case}. One clearly sees that $\work_S \geq \work_\Lambda$ for $\gamma \in [0,1]$. 
Note that, if Charlie does not perform the measurement on the control system $C$, i.e., $\Sup(\rho)=\tr_C\left[\rho_{C\W}\right]=\Lambda(\rho)$, one can observe that $\work_\Sup = \work_\Lambda$. This means that the operation conducted by the second demon, Charlie, is essential for enhancing the extracted work.
%in the presence of the pure dephasing noise.

\section{Circuit Realizations and noise simulations}\label{sec:exp_realiza}
In this section, we provide a circuit model for the two-demon heat engine with the superposition of two pure dephasing channels. In addition, we  implement the circuit on IBMQ  and IonQ quantum computers, where the enhancement due to the second demon, Charlie, can be clearly observed. Morevoer, we introduce the noise simulation algorithm and compare the results from IBMQ and IonQ devices with the noise simulations.
% Also, the experimental results agree with the noise model for the intrinsic errors of the device.

% \begin{figure}
%     \centering
%     \includegraphics[width=0.8\linewidth]{WorkExtraction/image/general circuit.png}
%     \caption{Experimental circuit for the two maxwell's demon engine. The superposed trajectory is realized by a series of controlled unitary operations.}
%     \label{fig:circuit}
% \end{figure}

\begin{figure*}[!htbp]
    \centering
    \includegraphics[width=1 \linewidth]{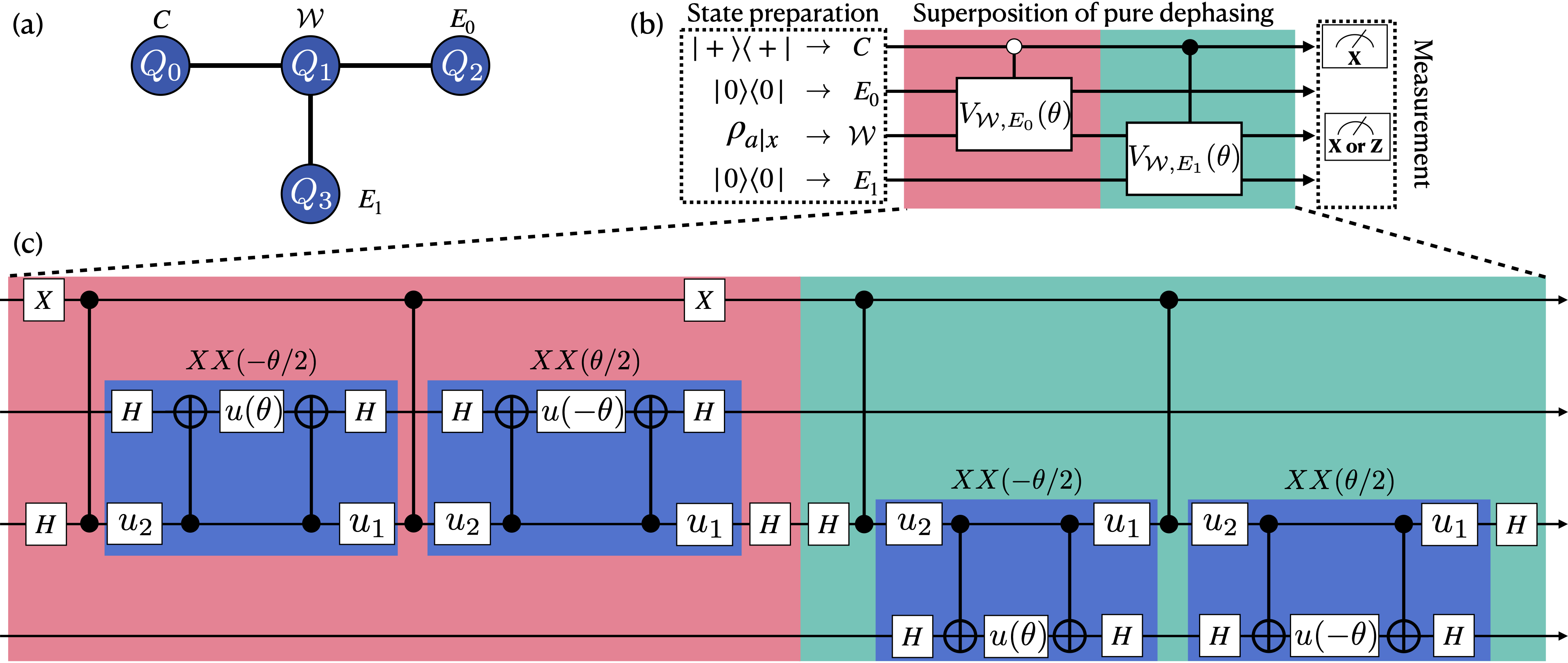}
    \caption{(a) The qubits we choose in \textit{ibmq\_jakarta} and the corresponding coupling map. Here, $Q_0$, $Q_1$, $Q_2$ and $Q_3$ serve as $C$,  $\W$, $E_0$ and $E_1$, respectively. (b) Circuit model for the temporal steering heat engine with superposition of dephasing channels. (c) The gate sequence for implementing the superposition of pure dephasing channels. Here, $u_1$, $u_2$, $u(\theta)$ are given by \eq{eq:u}.
    }
    \label{fig:whole_circuit}
\end{figure*}

%\subsection{Circuit model for the temporal steering heat engine and the experimental results}
The circuit model consists of four qubits: the control $C$, the work medium $\mathcal{W}$, and the environments, $E_0$ and $E_1$. The connections between qubits and the labels on IBMQ device are shown in \fig{fig:whole_circuit}(a). On the other hand, due to the full connectivity of the trapped ion device, we can choose four arbitrary qubits on IonQ device for the circuit.
%\blue{Since the trapped ion device doesn't need to consider the qubit connections, we can choose four arbitrary qubits on IonQ device for the circuit.}

The circuit model is described in \fig{fig:whole_circuit}(b) and can be divided into three parts: (1) state preparation, (2) superposition of pure dephasing channels, (3) measurements on $C$ and $\mathcal{W}$. In part (1), we prepare the control system $C$ and the environments in $\proj{+}$ and $\proj{0}$, respectively. Also, 
% since IBMQ does not allow one to access the post-measurement states \orange{HY: I remember IBM can do this now. Be careful.},
we replace Alice's measurements by directly preparing the post-measurement state $\rho_{a|x}$ in Table~\ref{tab:operators} and assign the probability $p(a|x)=1/2~\forall a, x$. %\orange{HY: do not mention artificial in the work otherwise the referee will think your work is ``artifiical". Unless this word provides postitive result.}

In part (2), we construct the $V_{\W,E}(\theta)$ according to \eq{eq:dephasing_model}, and the dephasing strength is modulated through the rotation angle $\theta$ with $\theta=\text{arccos}(1-\gamma)$. 
In addition, to achieve the superposition of quantum channels, we make a series of controlled unitaries according to \eq{eq:total_trans}. As shown in \fig{fig:whole_circuit}(c) (see also \apx{apx:circuit_decomp} for details), the controlled unitaries can be decomposed into controlled-$Z$ gate, and Ising coupling gate  $XX(\theta)=\cos\theta/2~\id \ts \id - i\sin\theta/2~\sigma_x \ts \sigma_x$, which is a native gate for IonQ trapped ion quantum computer.
% Since the basic two-qubit gate in IonQ device is Ising coupling gate $XX(\theta)=\cos\theta/2~\id \ts \id - \sin\theta/2~\sigma_x \ts \sigma_x$, we decompose the superposition of dephasing channels into $XX(-\theta/2)$, $XX(\theta/2)$, Controlled-$Z$, and $H$. 
Nevertheless, because Ising coupling gate is not a native gate in IBMQ,
we further decompose the $XX$ gate into CNOT gates, $H$ gates, and three other single-qubit gates listed as follows:
\begin{equation}\label{eq:u}
\begin{aligned}
    &u_1=\frac{1}{\sqrt{2}}
    \begin{pmatrix}
        1&-i\\
        -1&-i
    \end{pmatrix},
    u_2=\frac{1}{\sqrt{2}}
    \begin{pmatrix}
        1&-1\\
        i&i
    \end{pmatrix},\\
    &u(\theta)=\text{diag}(1,e^{i \theta /2}).  
\end{aligned} 
\end{equation}
 Note that, $u_1=U3(\pi/2, \pi,\pi/2)$, $u_2=U3(\pi/2, \pi/2,0)$, and $u(\theta)=P(\theta/2)$ can be easily implemented in IBMQ, where
\begin{equation}
\begin{aligned}
    P(\theta)&=
    \begin{pmatrix}
        1 & 0 \\
        0 & e^{i \theta}
    \end{pmatrix}  \\
    \mathrm{and~}U3(\theta, \phi, \xi)&=
    \begin{pmatrix}
        \cos{(\frac{\theta}{2})} & -e^{i\xi} \sin{(\frac{\theta}{2})}\\
        
        e^{i\phi}\sin{(\frac{\theta}{2})} & e^{i(\phi+\xi)}\cos{(\frac{\theta}{2})}
    \end{pmatrix}.
\end{aligned}
\end{equation}

%Here, $u_1$ and $u_2$ are constructed from the $U3(\alpha, \beta, \gamma)$ gate in IBM Q, where $u_1$ and $u_2$ are equal to $U3(\pi/2, \pi, \pi/2)$ and $U3(\pi/2, \pi/2, 0)$, respectively. The $u(\theta)$ is the rotation about Z-axis.

In part (3), we measure $C$ in $\sigma_x$ basis and denote $0$ ($1$) as the outcome associated with the eigenstate $\ket{+}_C$ ($\ket{-}_C$). For the single-channel scenario,
% where Charlie does not perform the selective measurement, 
both of the outcomes are taken into account; whereas, for the superposition scenario, 
% where Charlie conducts the selective measurement, 
we post-select the outcome $0$. Furthermore, as indicated in \eq{eq:avg_extracted_work}, the average extracted work can be obtained by the expectation values $\{\langle F_{a|x}\rangle= \tr(F_{a|x}\rho_{a|x})\}_{a,x}$. In addition, as summarized in \tab{tab:operators}, each $\langle F_{a|x} \rangle$ can be constructed by $\{\langle \sigma_i \rangle_{a|x} = \tr(\sigma_i \rho_{a|x})\}_{i \in \{x,z\}}$. Therefore, instead of applying the work extraction unitaries $\{U_{a|x}\}$, we measure the observables $\sigma_x$ and $\sigma_z$ on $\mathcal{W}$ to estimate $\work_{\Lambda}$ and $\work_{\Sup}$.

% \orange{[ 
% In (ii), you only ask the readers to see Fig. but how and why you can use that circuit to simulate the superposition channel? When I saw Fig. 3, I cannot understand too.]}

\begin{figure*}[!htbp]
    \centering
    \includegraphics[width=1\linewidth]{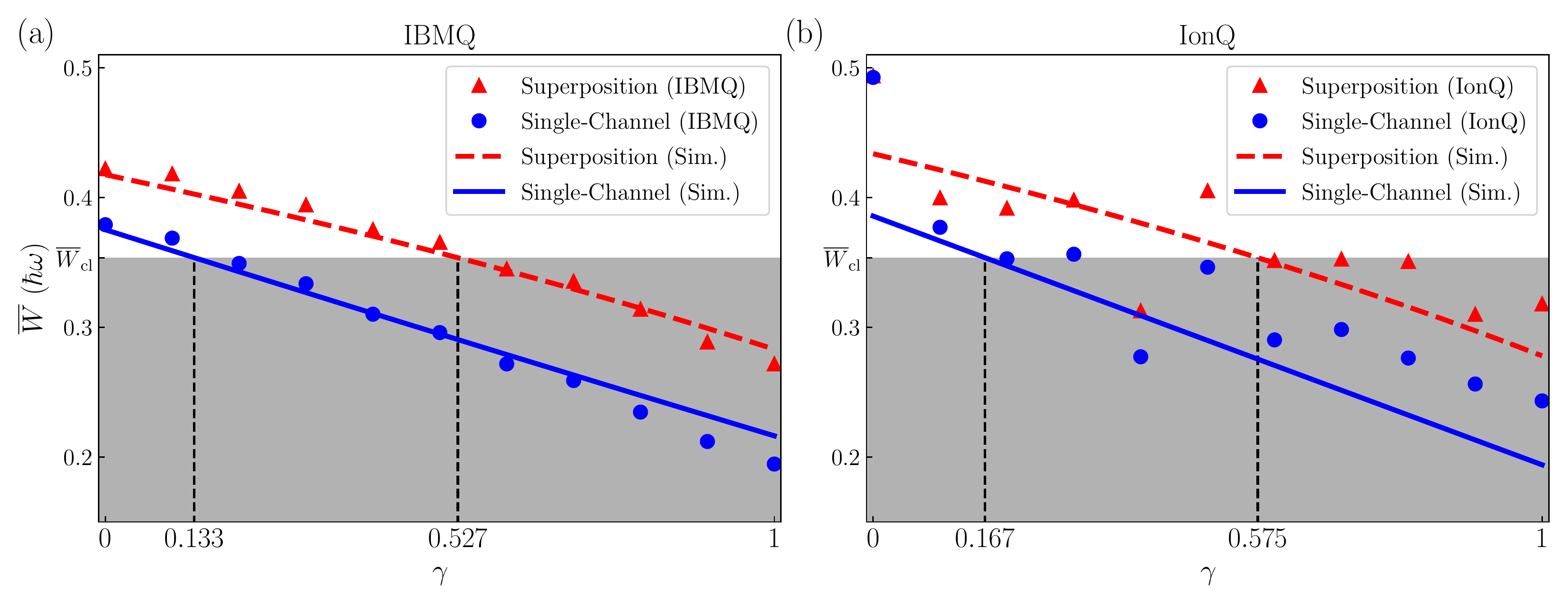}
    \caption{The noise simulations and results from (a) IBMQ and (b) IonQ devices. The results for $\work_{\Lambda}$ and $\work_{\Sup}$ are represented by blue dots and red triangles, respectively. The corresponding noise simulations are represented by blue-solid and red-dashed curves. Both the noise simulations and results from the devices show that $\work_\Sup > \work_\Lambda$.
    (a) For IBMQ, the results fit the noise simulations well. The results from the noise simulations show that $\work_\Sup$ and $\work_{\Lambda}$ reach the classical limit $\work_\cl$ at $\gamma \approx 0.133$ and $\gamma \approx 0.527$, respectively. (b) For IonQ, the results show an oscillatory deviation from the noise simulations because of the non-Markovian nature of the device's intrinsic errors. In addition, $\work_\Sup$ ($\work_{\Lambda}$) obtained from IonQ reaches the classical limit $\work_\cl$ at $\gamma \approx 0.167$ ($\gamma \approx 0.575$) and are both larger than those from IBMQ.
    }
    \label{fig:both_result}
\end{figure*}

In \fig{fig:both_result}, we present the results obtained from the devices of \textit{ibmq\_jakarta} and IonQ. For each circuit, it runs with $8192$ shots and $1000$ shots for IBMQ and IonQ devices, respectively.
% \red{[JD:It could be not so accurate because of the selection. Shall we also present the result for the successful counts?]} 
The blue dots and the red triangles represent the average extracted work for the scenarios with and without the post-selection, respectively. We can observe a clear enhancement due to the postselection. 
To gain a further insight on the devices' imperfections, we perform noise simulations~\cite{Ku2020, Huang2021} (the solid and dashed curves in \fig{fig:both_result}) that take the intrinsic errors of the quantum devices into account. To model the intrinsic errors, we consider three major sources of the errors, including the qubit relaxation, the qubit decoherence, the two-qubit gate error, and the readout error. The corresponding relaxation rate ($\eta_{T1}$), decoherence rate ($\eta_{T2}$), and the error rates of the devices, \textit{ibmq\_jakarta} and IonQ, are summarized in \tab{tab:errors1} and \tab{tab:errors2}. %As shown in \fig{fig:noise}, one can find that noise simulations agree with the experimental results. 
In the following, we elaborate on the noise model in detail. 

The qubit relaxation and decoherence are described by the following Lindblad master equation:
\begin{equation}\label{eq:mastereq}
\begin{aligned}
    \dot{\rho}(t)
    :=&\mathcal{L}\left[\rho(t)\right]\\
    =& \frac{\eta_{T_1}}{2} [ 2\sigma_- \rho(t) \sigma_+ - \sigma_+\sigma_-\rho(t)-\rho(t)\sigma_+\sigma_-]\\
    &+ \eta_{T_2}[\sigma_z\rho(t)\sigma_z-\rho(t)].
\end{aligned}
\end{equation}
When each ideal quantum gate (or unitary transformation) is applied to the system, we apply the propagation of the state according to \eq{eq:mastereq}, i.e., $\exp(\mathcal{L}t)$, where t is the gate time. %Note that, the average $\gamma_{T_1}$ and $\gamma_{T_2}$ are $10^{-4}~ (1/\text{ns})$ and $5 \times 10^{-3}~(1/\text{ns})$, respectively.

Among the gates we implement, two-qubit gates are the major source of gate errors because their gate time is almost one order of magnitude longer than that of single-qubit gates. Following the idea from Refs.~\cite{Magesan2010,Magesan2012,Urbanek2021prl}, we model the gate error as a depolarizing channel denoted as
% the CNOT error can then be modeled by a depolarizing channel, i.e.,
\begin{equation}\label{eq:CNOTErr}
\begin{aligned}
\mathcal{G}_{\text{er}}(\rho)= (1-p_{\text{GE}})\rho + p_{\text{GE}} \frac{\id}{2},
\end{aligned}
\end{equation}
where $p_{\text{GE}}$ is the two-qubit gate error rate. We apply the model after the Lindblad master equation only when a two-qubit gate is operated on the circuit.
% \blue{\st{Note that, among the errors we considered, the two-qubit gate errors are the most influential one on the experimental data.}}

To model the readout errors, we recall that quantum computers measure the systems in a computational basis, $\ket{0}$ and $\ket{1}$. The outcome is 0 (1) when the qubit is in $\ket{0}$ ($\ket{1}$) in ideal case. Hence, we can determine the readout error rate $\Gamma$ by the average probability of measuring $\ket{0}$ ($\ket{1}$) but obtaining the opposite outcome 1 (0). Therefore, the readout errors can be modeled as a bit-flip channel, i.e., 
\begin{equation}\label{eq:readerror}
    \mathcal{E}_{\text{readout}}(\rho) = (1-\Gamma) \rho + \Gamma~\sigma_x \rho \sigma_x.
\end{equation} 
% \red{[JD: how this errors accumulate in the noise model is not presented. Better to include this.]}

\begin{figure}[!htbp]
    \centering
    \includegraphics[width=1\linewidth]{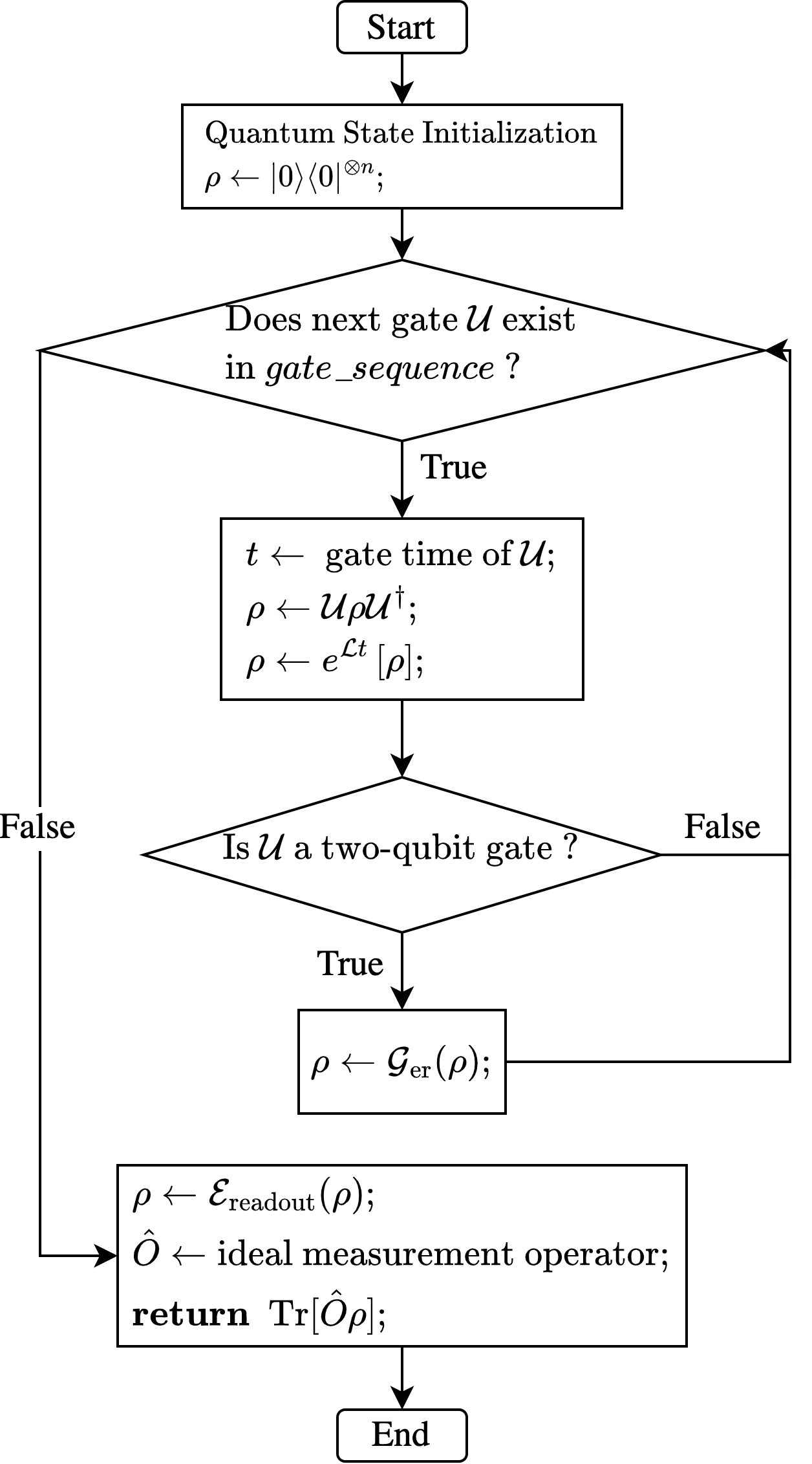}
    \caption{
    The flowchart of the noise simulation algorithm for a given $n-$qubit quantum circuit with a specific \textit{gate\_sequence}.
    }
    \label{fig:NS_flowchart}
\end{figure}

Here, we give a brief summary of our noise simulation process and how this is integrated with the superposition scenario in \sect{sec:superposition_of_quantum_channels}. We first modulate the dephasing strength $\gamma$ through the rotation angle $\theta$ with $\theta=\arccos(1-\gamma)$ and provide a circuit model [as shown in \fig{fig:whole_circuit}(c)] for the superposition scenario in \sect{sec:superposition_of_quantum_channels}. Thus, in order to obtain the average extracted work for a specific $\gamma$ through the circuit model, one needs to generate totally six quantum circuits because of the four different state preparations and two different measurement observables as presented in Table~\ref{tab:operators}. For each quantum circuit, one can apply the noise simulation by considering the intrinsic errors (as presented in Table~\ref{tab:errors1} and Table~\ref{tab:errors2}) including the qubit relaxation and decoherence in Eq.~(\ref{eq:mastereq}), the two-qubit gate error in Eq.~(\ref{eq:CNOTErr}), and the measurement error in Eq.~(\ref{eq:readerror}). The noise simulation algorithm for a given quantum circuit can be summarized by the flowchart presented in \fig{fig:NS_flowchart}.
% \begin{tcolorbox}[title=Noise Simulation Algorithm]
% \begin{algpseudocode}
% \State \textbf{Given} a quantum circuit (QC) which contains a \textit{gate\_sequence} with $n$ qubits.

% \textbf{Initialize} $\rho \leftarrow \proj{0}^{\otimes n}$;
% \For{\textbf{each} $\mathcal{U}$ \textbf{in} \textit{gate\_sequence}}

% \State $t \leftarrow$ gate time of $\mathcal{U}$;

% \State $\rho \leftarrow \mathcal{U} \rho \mathcal{U}^\dagger$; %\Comment{apply ideal gates}

% \State $\rho \leftarrow e^{\mathcal{J}t} [\rho]$; %\Comment{relaxation and dephasing error}

% \If{$\mathcal{U}$ \textbf{is} two-qubit gate}
% \State $\rho \leftarrow \mathcal{G}_{\text{er}}(\rho)$; %\Comment{two-qubit gate error}
% \EndIf
% \EndFor

% \State $\rho \leftarrow \mathcal{E}_\text{readout}(\rho)$; %\Comment{readout error}

% \State $\hat{O} \leftarrow$ ideal measurement operator in QC;

% \Return $\text{Tr}[\hat{O}\rho]$;
% \end{algpseudocode}
% \end{tcolorbox}

\begin{table}[!htbp]
  \caption{A summary of the intrinsic errors for the \textit{ibmq\_jakarta} and IonQ devices, which include the qubit relaxation rate $\eta_{T_1}$, decoherence rate $\eta_{T_2}$, and readout error rate $\Gamma$. The gate time for single-qubit gate are also presented in the table.}
  \label{tab:errors1}
  \begin{tabular}{cccccc}
      \hline\hline
      & \multirow{2}{*}{Qubit} & \multirow{2}{*}{$\eta_{T_1}(\mathrm{ns}^{-1})$} ~& \multirow{2}{*}{$\eta_{T_2}(\mathrm{ns}^{-1})$} 
      & \multirow{2}{*}{\shortstack{single-qubit\\gate time (ns)}} & \multirow{2}{*}{$\Gamma$} \\[1pt] 
      
      & & & &  & \\[1pt]
      \hline
      
      IBMQ
      & $Q_0$& $6.35 \times 10^{-6}$ & $1.45 \times 10^{-5}$ & 35.56 & $1.95 \%$ \\[1pt]
      
      & $Q_1$& $7.62 \times 10^{-6}$ & $2.78 \times 10^{-5}$ & 35.56 &  $2.12 \%$ \\[1pt]
      
      & $Q_2$& $6.94 \times 10^{-6}$ & $2.01 \times 10^{-5}$ & 35.56 &  N/A\\[1pt]
      
      & $Q_3$& $1.02 \times 10^{-5}$ & $1.44 \times 10^{-5}$ & 35.56 &  N/A\\[1pt]
      \hline
      IonQ
      & All & $10^{-10}$ & $2.48 \times 10^{-9}$ & $10^{4}$ & $0.39\%$ \\[1pt]
      
      \hline\hline
  \end{tabular}
%   \begin{tabular}{ccccc}
%       \hline\hline
      
%         & 0 & 1 & 2 & 3\\[1pt]
      
%       $\gamma_{T_1}^\text{IBM}~(1/\text{ns})$ & $6.35 \times 10^{-6}$ & $7.62 \times 10^{-6}$ & $6.94 \times 10^{-6}$ & $1.02 \times 10^{-5}$\\[1pt]
      
%       $\gamma_{T_2}^\text{IBM}~(1/\text{ns})$ & $1.45 \times 10^{-5}$ & $2.78 \times 10^{-5}$ & $2.01 \times 10^{-5}$ & $1.44 \times 10^{-5}$ \\[1pt]
      
%       $\gamma_{T_1}^\text{IonQ}~(1/\text{ns})$ & \multicolumn{4}{c}{$10^{-4}$} \\[3pt]
      
%       $\gamma_{T_2}^\text{IonQ}~(1/\text{ns})$ & \multicolumn{4}{c}{$5 \times 10^{-3}$} \\[3pt]
      
%       \hline
      
%       $p_{CN_{0,1}}$ & \multicolumn{4}{c}{$6.7 \times 10 ^{-3}$}  \\[3pt]
      
%       $p_{CN_{1,2}}$ & \multicolumn{4}{c}{$1.2 \times 10^{-3}$}  \\[3pt]
      
%       $p_{CN_{1,3}}$ & \multicolumn{4}{c}{$8 \times 10^{-3}$} \\[3pt]
      
%       $p_{XX}$         & \multicolumn{4}{c}{$3.04 \times 10^{-2}$} \\[3pt] 
%       %\cmidrule(l){2-5}
      
%       \hline
      
%       $\Gamma^\text{IBM}$ & 0.020 & 0.021 &  &  \\[1pt]
      
%       $\Gamma^\text{IonQ}$ & 0.0039 & 0.0039 & & \\[1pt]
      
%       \hline\hline
%   \end{tabular}
\end{table}

\begin{table}[!htbp]
  \caption{A summary of the two-qubit gate time and two-qubit gate error $p_{\text{GE}}$ for the \textit{ibmq\_jakarta} and IonQ devices. The CNOT (CN) gate is the two-qubit gate for IBMQ devices and the subscripts represent the qubit index [see Fig.~\ref{fig:whole_circuit}(a)] of which the CN gate are applied. The Ising coupling gate ($XX$) is the two-qubit gate for IonQ devices. In addition, due to the full connectivity of the trapped ion device, we can apply $XX$ gate to two arbitrary qubits.}
  \label{tab:errors2}
  \begin{tabular}{cccc}
      \hline\hline
      & \multirow{2}{*}{~Gate~}
      & ~\multirow{2}{*}{\shortstack{two-qubit\\gate time (ns)}}~
      & \multirow{2}{*}{$p_\mathrm{GE}$} \\[1pt] 
      
      & & & \\[1pt]
      \hline
      
      IBMQ
      & $\mathrm{CN_{0,1}}$ & 234.67 & $0.67 \%$ \\[1pt]
      
      & $\mathrm{CN_{1,2}}$ & 284.44 &  $0.12 \%$ \\[1pt]
      
      & $\mathrm{CN_{1,3}}$ & 384.00 & $0.8 \%$ \\ [1pt] 
      \hline
      IonQ
      & ${XX}$ & $2.1 \times 10^{5}$ & $3.04\%$ \\[1pt]
      
      \hline\hline
  \end{tabular}
\end{table}

% \begin{figure}
%     \centering
%     \includegraphics[width=0.98 \linewidth]{noise_simulation_and_experiment_result.pdf}
%     \caption{The results from IBMQ device and noise simulations. The experimental results for $\work_{\Lambda}$ and $\work_{\Sup}$ are represented by blue dots and red triangles, respectively. The corresponding noise simulations are represented by blue-solid and red-dashed curves. Here, $\gamma_{\text{TH},\text{Noise}}\approx 0.17$ and  $\gamma_{\Sup,\text{Noise}}\approx 0.56$ are the corresponding quantum-to-classical transition thresholds predicted by noise simulations.}
%     \label{fig:noise}
% \end{figure}

As shown in \fig{fig:both_result}, one can observe that the results from the IBMQ agree with the noise model. However, the results from IonQ do not fit the simulations well. The result at $\gamma=0$ is nearly the same as the ideal result, instead of that from noise simulations. It is because the IonQ software provides compulsorily optimized gates, and the total unitary of the circuit at $\gamma=0$ is effectively the same as $\id$ applied to four qubits. 
% \blue{\st{Moreover, the IonQ results also show oscillations as $\gamma$ increases. Since the IonQ device uses laser beams to couple all of the qubits and focus on each ion, it may be the reason for the oscillations.}}

In order to better understand how each intrinsic error affects the work medium and, in particular, the value of average extracted work $\overline{W}$, we simplify the discussion by only considering the errors on the work medium ($Q_1$) and taking dephasing strength $\gamma=0$ under single-channel scenario as an example. We first discuss the effect of single-qubit relaxation and decoherence errors on the work medium. As shown in Fig.~\ref{fig:whole_circuit}(c), the circuit has totally 26 (6) single-qubit gates and 12 (8) two-qubit gates in IBMQ (IonQ) device. Because the controlled-$Z$ operation is further decomposed into one CNOT gate and two Hadamard gates, one can then estimate the total gate time for IBMQ and IonQ devices are approximately $4.82\times10^3$ and $1.98\times10^6~\text{ns}$, respectively. By solving \eq{eq:mastereq}, one can further calculate $\overline{W}$ by only considering the qubit relaxation and decoherence error on the work medium ($Q_1$). The value obtained for IBMQ and IonQ devices are approximately $0.4107 \hbar \omega$ and $0.4974 \hbar \omega$, respectively. In this case, the ideal value of $\work$ is $0.5\hbar \omega$, and one can thus define the error rate of qubit relaxation and decoherence for both devices, namely $p_{T_1, T_2}^{\text{IBMQ}}=17.86\%$ and $p_{T_1, T_2}^{\text{IonQ}}=0.52\%$. Next, we discuss the influence of the two-qubit gate error on both devices. For the IonQ device, since there are 8 two-qubit gates applied to the work medium $Q_1$ in total, the effective two-qubit gate error rate for the IonQ device is given as $p_{\text{GE}}^{\text{IonQ}}=1-(1-3.04\%)^8\approx21.88\%$. For the IBMQ device, since each CNOT gate ($\text{CN}_{0,1}$, $\text{CN}_{1,2}$, and $\text{CN}_{1,3}$) is applied four times to the work medium $Q_1$, the effectively two-qubit gate error rate for IBMQ device can be given as $p_{\text{GE}}^{\text{IBMQ}}=1-[(1-0.67\%)(1-0.12\%)(1-0.8\%)]^4\approx6.18\%$. 
Combining all the error rates including qubit relaxation, qubit decoherence, two-qubit gate, and readout, the total error rates for IBMQ and IonQ devices are given by 
\begin{align}
    & 1-(1-p_{T_1, T_2}^{\text{IBMQ}})(1-p_{\text{GE}}^{\text{IBMQ}})(1-\Gamma^{\text{IBMQ}}_{Q_1})\approx24.57\%\notag \\
    \text{and} ~~& 1-(1-p_{T_1, T_2}^{\text{IonQ}})(1-p_{\text{GE}}^{\text{IonQ}})(1-\Gamma^{\text{IonQ}})\approx22.59\%,\notag
\end{align}
respectively. As seen, the total error rates for both devices are quite similar, which can also be observed from \fig{fig:both_result} (the noise simulation for $\gamma=0$ under single-channel scenario). Note that, for real noise simulation, one has to further take the errors on the other three qubits ($Q_0$, $Q_2$, and $Q_3$) into account.
%On the other hand, the noise simulations results of IBMQ and IonQ are similar. As shown in \fig{fig:both_result}, the quantum-to-classical transition thresholds of the single-channel and superposition scenario for IBMQ (IonQ) are approximately 0.164 (0.167) and 0.550 (0.575), respectively. We first discuss the qubit relaxation and decoherence effect for this case. From the gate times summarized in Table~\ref{tab:errors1} and Table~\ref{tab:errors2}, one can calculate the total gate time of the circuit in \fig{fig:whole_circuit}(c) is approximately $4 \times 10^3$~ns for IBMQ device and $2.6 \times 10^6$~ns for IonQ device. For the qubit relaxation rate $\eta_{T1}$ and decoherence rate $\eta_{T2}$ presented in Table~\ref{tab:errors1}, one can observe that for both devices, the total gate time are short so the relaxation and decoherence error are not significant for this circuit design. The main difference between the errors of the devices are the two-qubit gate error $p_{\text{GE}}$ and the readout error $\Gamma$ presented in Table~\ref{tab:errors2}. Although the average value of two-qubit gate error for IonQ is about six times larger than IBMQ's, the average value of readout error for IBMQ is about six times larger than IonQ's. From numerical results, we found that noise simulations of IonQ and IBMQ devices are similar because the effective total error cause by two-qubit gate and readout are quite close.

The oscillations of the results obtained from IonQ device, as shown in \fig{fig:both_result}(b), could originate from the non-Markovian nature of the device's intrinsic errors. More specifically, in our noise simulations, we implicitly assume that the errors of each gate are independent of each other (i.e., the model respects Markov approximation). That is, for the circuits with the same depth (e.g., the circuit model for different dephasing strength $\gamma$), the differences between the ideal results and the noise simulations are roughly the same. Therefore, we don't observe oscillations from the noise simulations in \fig{fig:both_result}(b). Therefore, the oscillations observed from the data in \fig{fig:both_result}(b) suggest that there could be some non-Markovian effects from the intrinsic errors in the IonQ device. For instance, as reported in Ref.~\cite{Romero2021prx}, the noise of each gate could be correlated and induces the non-Markovian effects.

% \begin{figure}
%     \centering
%     \includegraphics[width=0.98 \linewidth]{IonQ_dephasing.pdf}
%     \caption{The experimental data obtains from IonQ also shows that the post-selected data can show the larger average extracted work.}
%     \label{fig:IonQ_exp}
% \end{figure}

\section{Conclusion}
In this work, we introduced the notion of a two-demon steering heat engine and obtained the corresponding classical limit for extractable work via SDP. We considered that the work medium passes through a pure dephasing channel. We observed a monotonic decrease of the extractable work and a threshold for quantum-to-classical transition. Further, we utilized a superposition of the pure dephasing channels and introduced the second demon. We can observe clear enhancements of the extractable work and the quantum-to-classical thresholds. Moreover, we performed proof-of-principle demonstrations on IBMQ and IonQ quantum computers. The results also demonstrate enhancements with the assistance of the second demon, Charlie. The results from IBMQ also agree with the noise simulations that include the accumulation of errors (qubit relaxation, qubit decoherence, two-qubit gate error, and readout error) during the process. On the other hand, the results from IonQ demonstrate an unexpected oscillated behavior, implying an intrinsic non-Markovian nature of the device.

The quantum channels considered in our work are pure dephasing. It will be interesting to generalize the present work to the cases of different channels, such as depolarizing channel or amplitude damping channel. In addition, the quantum control can also be used in the case of indefinite causal order~\cite{Ebler2018,Loizeau2020,Felce2020,Guha2020,Simonov2022}. How the second demon affect the extractable work in this case also deserves further investigations.

\section{acknowledgement}
% We acknowledge the NTU-IBM Q Hub (Grant: MOST 107-2627-E-002-001-MY3), the IBM quantum experience and IonQ for providing us a platform to implement the experiment. The views expressed are those of the authors and do not reflect the official policy or position of IBM or the IBM Quantum Experience team.
% The authors acknowledge the support from the National Center for Theoretical Sciences and Ministry of Science and Technology, Taiwan (Grants Nos. MOST 107-2628-M-006-002-MY3, 109-2627-M-006-004), the National Center for Theoretical Sciences and Ministry of Science and Technology, Taiwan (Grant No. MOST 108-2811-M-006-536) for H.-Y.K., and Army Research Office (Grant No. W911NF-19-1-0081) for Y.-N.C.
%The authors acknowledge fruitful discussions with Yi-Te Huang, Jhen-Dong Lin, and Huan-Yu Ku.
We acknowledge the NTU-IBM Q Hub and Cloud Computing Center for Quantum Science \& Technology at NCKU for providing us platforms to implement the circuits.
The views expressed are those of the authors and do not reflect the official policy or position of IBM or the IBM Quantum Experience team.
H.-Y. K. is supported by the National Center for Theoretical Sciences and National Science and Technology Council, Taiwan (Grant No. MOST 110-2811-M-006-546 and No. MOST 111-2917-I-564-005). H.-B.C. is supported by the National Science and Technology Council, Taiwan, Grant No. MOST 110-2112-M-006-012. This work is supported by the National Center for Theoretical Sciences and National Science and Technology Council, Taiwan, Grant No. MOST 111-2123-M-006-001, and Grant No. Most 110-2627-M-006-004.

\appendix

\section{Semidefinite program for the classical limits of average extracted work}\label{apx:classical_bound}
In this section, we briefly summarize how to obtain the classical limit of the average
extracted work by introducing the hidden-state (HS) model and semidefinite program (SDP).

Bob will conceive that the assemblage $\{p(a|x)\Lambda(\rho_{a|x})\}$ is classical whenever it can be decomposed with HS model~\cite{Wiseman2007} as Eq.~(\ref{eq:HS}). If Bob can
construct any one of such decompositions, he considers that the extracted work is produced
by classical resources. Therefore, for a given work-extraction protocol, consisting of $\lbrace F_{a|x} \rbrace_{a|x}$ and the Gibbs state $\rho_\W$, the classical limit of average extracted work is the maximal value attainable by classical assemblages (described by HS model).
Besides, since all the marginals of HS model should result in the same Gibbs state $\rho_\W$, as discussed in \sect{sec:temporal_steering_heat_engine}, this imposes an additional constraint on the definition of the classical limit. Furthermore, because the set of all HS models (denote as $\mathcal{C}$) is a convex set, we can define the classical limit of average extracted work as a convex optimization problem, recalled from the main context,
\begin{align}\label{eq:Wcl_convex_opt}
\work_\cl=&\max_{\assemb{\tilde{\sigma}_{a|x}}\in\mathcal{C}}~\sum_{a,x}\tr\left[F_{a|x} \tilde{\sigma}_{a|x}\right],\nonumber\\
&\text{s.t.}~\sum_a \tilde{\sigma}_{a|x}=\rho_\W~\forall~x.
\end{align}

%The LHS model~\cite{Wiseman2007, Jones2007} conceives an assemblage classical whenever it can be decomposed with a set of local hidden states as Eq.~(\ref{eq:LHS}). Consequently, the classical average extracted work means that the assemblage $\{p(a|x)\Lambda(\rho_{a|x})\}$ in \eq{eq:avg_extracted_work} can be described by Eq.~(\ref{eq:LHS}). We denote the set of all LHS models as $\mathcal{L}$, which is a convex set.

To solve the above optimization problem, we recast the HS model by introducing the deterministic strategy~\cite{Uola2020, Cavalcanti2016}. Let us consider $x\in\assemb{1,2,...,m}$ and $a\in\assemb{1,2,...,q}$. Since $m$ and $q$ are finite, the number of hidden variable $\lambda$ is $q^m$. Each $\lambda$ can be considered as a string of ordered outcomes according to the measurements, i.e., $\lambda=(a_{x=1}, a_{x=2}, ... ,a_{x=m})$. We can now define the deterministic strategy with $\delta_{a, \lambda(x)}$, where $\delta$ is the Kronecker delta function and $\lambda(x)$ denotes the value of $a$ at position $x$. Therefore, the HS model in \eq{eq:HS} can be expressed by the deterministic strategy $\delta_{a,\lambda(x)}$ together with a set of unnormalized hidden states $\{\sigma_\lambda\}_\lambda$, namely $\sigma_{a|x}^{\text{HS}}=\sum_\lambda\delta_{a,\lambda(x)}\sigma_\lambda$. We can then recast the constraint in \eq{eq:Wcl_convex_opt} as
\begin{alignat}{1}
    &\sum_a \tilde{\sigma}_{a|x}=\sum_{a, \lambda} \delta_{a, \lambda(x)}\sigma_\lambda=\sum_\lambda \sigma_\lambda=\rho_\W~~\forall~x\notag\\
    &\text{and}~\sigma_\lambda \geq 0~~\forall~~\lambda.
\end{alignat}
The second equality in the first constraint holds because $\sum_a \delta_{a, \lambda(x)}=1$ for all $\lambda$ and $x$. Since $\{\sigma_\lambda\}_\lambda$ belongs to convex set, and all the constraints are linear, one can obtain the optimal solution of \eq{eq:Wcl_convex_opt} by solving the following semidefinite program: 
\begin{alignat}{1}\label{eq:Wcl_sdp}
    \work_\cl = \max_{\assemb{\sigma_{\lambda}}}&\sum_{a,x,\lambda}\delta_{a, \lambda(x)}\tr\left[F_{a|x} \sigma_{\lambda}\right],\notag\\
    \text{s.t.}&~\sum_\lambda \sigma_\lambda = \rho_\W,\notag\\
    &~\sigma_\lambda \geq 0 ~\forall~ \lambda.
\end{alignat}
Finally, we obtain the value of the classical limit of average extracted work.

Here, we take the case summarized in Table~\ref{tab:operators} as an example and provide the optimal solution obtained from the SDP in \eq{eq:Wcl_sdp}. Since $x\in\{\sigma_x, \sigma_z\}$ and $a\in\{+1, -1\}$, the number of local hidden variable $\lambda$ is four, and each $\lambda=(a_{x=\sigma_z}, a_{x=\sigma_x})$. Therefore, by numerically solving the SDP in Eq.~(A3), one can obtain the optimal value of $\overline{W}_{\text{cl}}\approx0.354\hbar\omega$, where the hidden variable $\lambda$ and the optimal solution $\{\sigma^*_\lambda\}$ can be summarized in Table~\ref{tab:sdp_opt_sol}. Furthermore, one can obtain the optimal HS assemblage:
\begin{alignat}{1}
&\sigma^{\text{HS}}_{a=+1|x=\sigma_z}\approx
        \begin{pmatrix}
            0.4268 &        0.0000\\
                   0.0000 & 0.0732
        \end{pmatrix},\notag\\
&\sigma^{\text{HS}}_{a=-1|x=\sigma_z}\approx
        \begin{pmatrix}
            0.0732 &         0.0000\\
                    0.0000 & 0.4268
        \end{pmatrix},\notag\\
&\sigma^{\text{HS}}_{a=+1|x=\sigma_x}\approx
        \begin{pmatrix}
                 0.2500 & 0.1768\\
            0.1768 &      0.2500
        \end{pmatrix},\notag\\
&\sigma^{\text{HS}}_{a=-1|x=\sigma_x}\approx
        \begin{pmatrix}
                  0.2500 & -0.1768\\
            -0.1768 &       0.2500
        \end{pmatrix}.
\end{alignat}
Note that the optimal solution $\{\sigma^*_\lambda\}$ is not unique, and  the above HS assemblage is therefore not unique either.

\begin{table}[!htbp]
    \caption{One of the optimal solution of \eq{eq:Wcl_sdp} for the case summarized in Table~\ref{tab:operators}. Each $\lambda$ can be considered as a string of ordered outcomes according to the measurements. Here, $\sigma_\lambda^*$ denotes the optimal solution for the unnormalized hidden states obtained from the SDP.}
    \label{tab:sdp_opt_sol}
\begin{tabular}{cc}
  \hline\hline
  $\lambda=(a_{x=\sigma_z}$, $a_{x=\sigma_x})$ & $\sigma^*_\lambda$\\
  \hline
  $(+1, +1)$ & 
      $\begin{pmatrix}
        0.2134 & 0.0884\\
        0.0884 & 0.0366
      \end{pmatrix}$\\
  $(+1, -1)$ & 
      $\begin{pmatrix}
        0.2134 & -0.0884\\
        -0.0884 & 0.0366
      \end{pmatrix}$\\
  $(-1, +1)$ & 
      $\begin{pmatrix}
        0.0366 & 0.0884\\
        0.0.0884 & 0.2134
      \end{pmatrix}$\\
  $(-1, -1)$ & 
      $\begin{pmatrix}
        0.0366 & -0.0884\\
        -0.0884 & 0.2134
      \end{pmatrix}$\\
  
  \hline\hline
\end{tabular}
\end{table}
\begin{figure*}
    \centering
    \includegraphics[width=0.98 \linewidth]{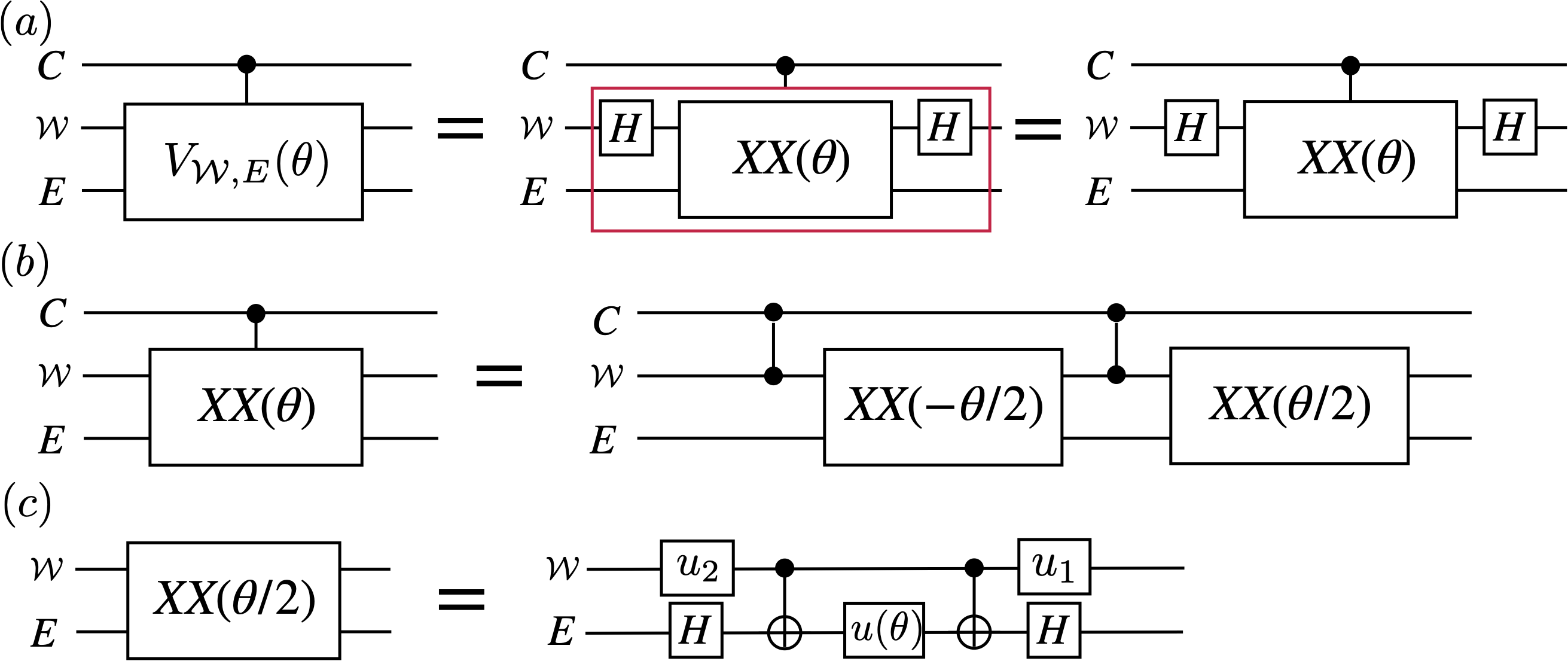}
    \caption{(a) The controlled-$V_{\W,E}$ can be decomposed into two $H$ gates and a controlled $XX(\theta)$ gate. (b) Controlled-$XX(\theta)$ can be decomposed to the combination of two controlled-$Z$ gates, $XX(-\theta/2)$ and $XX(\theta/2)$. (c) The decomposition of $XX(\theta)$ are $u_1$, $u_2$, $u(\theta)$, two $H$ gates, two CNOT gates and $H$ gates.}
    \label{fig:circuit_decomp}
\end{figure*}
\section{The decomposition of controlled-$V_{\W,E}$ operations}\label{apx:circuit_decomp}

In this section, we describe the detail how we decompose the controlled-$V_{\W,E}$ into the combination of two-qubit gates and single-qubit gates that can be applied on IBMQ and IonQ platforms.

According to \eq{eq:dephasing_model}, we can construst the $V_{\W,E}(\theta)$ as
\begin{equation}\label{eq:Vw}
    V_{\W,E}(\theta)=
    \begin{pmatrix}
        \cos(\frac{\theta}{2}) & -i\sin(\frac{\theta}{2}) & 0 & 0\\
        -i\sin(\frac{\theta}{2}) & \cos(\frac{\theta}{2}) & 0 & 0\\
        0 & 0 & \cos(\frac{\theta}{2}) & i\sin(\frac{\theta}{2}) \\
        0 & 0 & i\sin(\frac{\theta}{2}) & \cos(\frac{\theta}{2})
    \end{pmatrix}.
\end{equation}
In addition, to achieve the superposition of quantum channels, we use a controlled unitary according to \eq{eq:total_trans}. Moreover, the controlled unitary is equal to the product of two controlled unitaries.

Now, we consider the controlled unitary: $\proj{0}_C\ts\id+\proj{1}_C\ts V_{\W,E}$, as shown in \fig{fig:circuit_decomp}(a). From \eq{eq:Vw}, $V_{\W,E}$ is equal to $(H\ts \id) XX(\theta)(H\ts \id) $, where $XX(\theta)$ is a kind of Ising coupling gate denoted as: $XX(\theta)=\cos(\theta/2)\id \ts \id - i\sin(\theta/2) \sigma_x \ts \sigma_x$. The controlled-$V_{\W,E}$ can be separated to three controlled unitaries: two controlled-$H$ operations and a controlled-$XX(\theta)$ operation. In this case, the two controlled-$H$ operations can be replaced with two $H$ gates applied to the $\W$. Thus, we can focus on the $XX(\theta)$. As shown in \fig{fig:circuit_decomp}(b), the controlled-$XX(\theta)$ can be decomposed to two controlled-$Z$ operations, $XX(-\theta/2)$, and $XX(\theta/2)$.

However, IBMQ doesn't have the Ising coupling gates. According to IBMQ, $XX(\theta)$ can be decomposed to $u_1$, $u_2$, $u(\theta)$, two $H$s and two CNOT operations. In \fig{fig:circuit_decomp}(c), we show the order of the aforementioned  unitary operations to construct the $XX(\theta)$.


\begin{thebibliography}{81}%
\makeatletter
\providecommand \@ifxundefined [1]{%
 \@ifx{#1\undefined}
}%
\providecommand \@ifnum [1]{%
 \ifnum #1\expandafter \@firstoftwo
 \else \expandafter \@secondoftwo
 \fi
}%
\providecommand \@ifx [1]{%
 \ifx #1\expandafter \@firstoftwo
 \else \expandafter \@secondoftwo
 \fi
}%
\providecommand \natexlab [1]{#1}%
\providecommand \enquote  [1]{``#1''}%
\providecommand \bibnamefont  [1]{#1}%
\providecommand \bibfnamefont [1]{#1}%
\providecommand \citenamefont [1]{#1}%
\providecommand \href@noop [0]{\@secondoftwo}%
\providecommand \href [0]{\begingroup \@sanitize@url \@href}%
\providecommand \@href[1]{\@@startlink{#1}\@@href}%
\providecommand \@@href[1]{\endgroup#1\@@endlink}%
\providecommand \@sanitize@url [0]{\catcode `\\12\catcode `\$12\catcode
  `\&12\catcode `\#12\catcode `\^12\catcode `\_12\catcode `\%12\relax}%
\providecommand \@@startlink[1]{}%
\providecommand \@@endlink[0]{}%
\providecommand \url  [0]{\begingroup\@sanitize@url \@url }%
\providecommand \@url [1]{\endgroup\@href {#1}{\urlprefix }}%
\providecommand \urlprefix  [0]{URL }%
\providecommand \Eprint [0]{\href }%
\providecommand \doibase [0]{https://doi.org/}%
\providecommand \selectlanguage [0]{\@gobble}%
\providecommand \bibinfo  [0]{\@secondoftwo}%
\providecommand \bibfield  [0]{\@secondoftwo}%
\providecommand \translation [1]{[#1]}%
\providecommand \BibitemOpen [0]{}%
\providecommand \bibitemStop [0]{}%
\providecommand \bibitemNoStop [0]{.\EOS\space}%
\providecommand \EOS [0]{\spacefactor3000\relax}%
\providecommand \BibitemShut  [1]{\csname bibitem#1\endcsname}%
\let\auto@bib@innerbib\@empty
%</preamble>
\bibitem [{\citenamefont {Scully}\ \emph {et~al.}(2011)\citenamefont {Scully},
  \citenamefont {Chapin}, \citenamefont {Dorfman}, \citenamefont {Kim},\ and\
  \citenamefont {Svidzinsky}}]{Scully2011}%
  \BibitemOpen
  \bibfield  {author} {\bibinfo {author} {\bibfnamefont {M.~O.}\ \bibnamefont
  {Scully}}, \bibinfo {author} {\bibfnamefont {K.~R.}\ \bibnamefont {Chapin}},
  \bibinfo {author} {\bibfnamefont {K.~E.}\ \bibnamefont {Dorfman}}, \bibinfo
  {author} {\bibfnamefont {M.~B.}\ \bibnamefont {Kim}},\ and\ \bibinfo {author}
  {\bibfnamefont {A.}~\bibnamefont {Svidzinsky}},\ }\bibfield  {title}
  {\bibinfo {title} {Quantum heat engine power can be increased by
  noise-induced coherence},\ }\href {https://doi.org/10.1073/pnas.1110234108}
  {\bibfield  {journal} {\bibinfo  {journal} {Proc. Natl. Acad. Sci. U.S.A.}\
  }\textbf {\bibinfo {volume} {108}},\ \bibinfo {pages} {15097} (\bibinfo
  {year} {2011})}\BibitemShut {NoStop}%
\bibitem [{\citenamefont {Rahav}\ \emph {et~al.}(2012)\citenamefont {Rahav},
  \citenamefont {Harbola},\ and\ \citenamefont {Mukamel}}]{Rahav2012}%
  \BibitemOpen
  \bibfield  {author} {\bibinfo {author} {\bibfnamefont {S.}~\bibnamefont
  {Rahav}}, \bibinfo {author} {\bibfnamefont {U.}~\bibnamefont {Harbola}},\
  and\ \bibinfo {author} {\bibfnamefont {S.}~\bibnamefont {Mukamel}},\
  }\bibfield  {title} {\bibinfo {title} {Heat fluctuations and coherences in a
  quantum heat engine},\ }\href {https://doi.org/10.1103/PhysRevA.86.043843}
  {\bibfield  {journal} {\bibinfo  {journal} {Phys. Rev. A}\ }\textbf {\bibinfo
  {volume} {86}},\ \bibinfo {pages} {043843} (\bibinfo {year}
  {2012})}\BibitemShut {NoStop}%
\bibitem [{\citenamefont {Brunner}\ \emph {et~al.}(2014)\citenamefont
  {Brunner}, \citenamefont {Huber}, \citenamefont {Linden}, \citenamefont
  {Popescu}, \citenamefont {Silva},\ and\ \citenamefont
  {Skrzypczyk}}]{Brunner2014}%
  \BibitemOpen
  \bibfield  {author} {\bibinfo {author} {\bibfnamefont {N.}~\bibnamefont
  {Brunner}}, \bibinfo {author} {\bibfnamefont {M.}~\bibnamefont {Huber}},
  \bibinfo {author} {\bibfnamefont {N.}~\bibnamefont {Linden}}, \bibinfo
  {author} {\bibfnamefont {S.}~\bibnamefont {Popescu}}, \bibinfo {author}
  {\bibfnamefont {R.}~\bibnamefont {Silva}},\ and\ \bibinfo {author}
  {\bibfnamefont {P.}~\bibnamefont {Skrzypczyk}},\ }\bibfield  {title}
  {\bibinfo {title} {Entanglement enhances cooling in microscopic quantum
  refrigerators},\ }\href {https://doi.org/10.1103/PhysRevE.89.032115}
  {\bibfield  {journal} {\bibinfo  {journal} {Phys. Rev. E}\ }\textbf {\bibinfo
  {volume} {89}},\ \bibinfo {pages} {032115} (\bibinfo {year}
  {2014})}\BibitemShut {NoStop}%
\bibitem [{\citenamefont {Mitchison}\ \emph {et~al.}(2015)\citenamefont
  {Mitchison}, \citenamefont {Woods}, \citenamefont {Prior},\ and\
  \citenamefont {Huber}}]{Mitchison2015}%
  \BibitemOpen
  \bibfield  {author} {\bibinfo {author} {\bibfnamefont {M.~T.}\ \bibnamefont
  {Mitchison}}, \bibinfo {author} {\bibfnamefont {M.~P.}\ \bibnamefont
  {Woods}}, \bibinfo {author} {\bibfnamefont {J.}~\bibnamefont {Prior}},\ and\
  \bibinfo {author} {\bibfnamefont {M.}~\bibnamefont {Huber}},\ }\bibfield
  {title} {\bibinfo {title} {Coherence-assisted single-shot cooling by quantum
  absorption refrigerators},\ }\href
  {https://doi.org/10.1088/1367-2630/17/11/115013} {\bibfield  {journal}
  {\bibinfo  {journal} {New J. Phys.}\ }\textbf {\bibinfo {volume} {17}},\
  \bibinfo {pages} {115013} (\bibinfo {year} {2015})}\BibitemShut {NoStop}%
\bibitem [{\citenamefont {Chen}\ \emph
  {et~al.}(2016{\natexlab{a}})\citenamefont {Chen}, \citenamefont {Chiu},\ and\
  \citenamefont {Chen}}]{HBChen2016}%
  \BibitemOpen
  \bibfield  {author} {\bibinfo {author} {\bibfnamefont {H.-B.}\ \bibnamefont
  {Chen}}, \bibinfo {author} {\bibfnamefont {P.-Y.}\ \bibnamefont {Chiu}},\
  and\ \bibinfo {author} {\bibfnamefont {Y.-N.}\ \bibnamefont {Chen}},\
  }\bibfield  {title} {\bibinfo {title} {Vibration-induced coherence
  enhancement of the performance of a biological quantum heat engine},\ }\href
  {https://doi.org/10.1103/PhysRevE.94.052101} {\bibfield  {journal} {\bibinfo
  {journal} {Phys. Rev. E}\ }\textbf {\bibinfo {volume} {94}},\ \bibinfo
  {pages} {052101} (\bibinfo {year} {2016}{\natexlab{a}})}\BibitemShut
  {NoStop}%
\bibitem [{\citenamefont {Brandner}\ \emph {et~al.}(2017)\citenamefont
  {Brandner}, \citenamefont {Bauer},\ and\ \citenamefont
  {Seifert}}]{Brandner2017}%
  \BibitemOpen
  \bibfield  {author} {\bibinfo {author} {\bibfnamefont {K.}~\bibnamefont
  {Brandner}}, \bibinfo {author} {\bibfnamefont {M.}~\bibnamefont {Bauer}},\
  and\ \bibinfo {author} {\bibfnamefont {U.}~\bibnamefont {Seifert}},\
  }\bibfield  {title} {\bibinfo {title} {Universal coherence-induced power
  losses of quantum heat engines in linear response},\ }\href
  {https://doi.org/10.1103/PhysRevLett.119.170602} {\bibfield  {journal}
  {\bibinfo  {journal} {Phys. Rev. Lett.}\ }\textbf {\bibinfo {volume} {119}},\
  \bibinfo {pages} {170602} (\bibinfo {year} {2017})}\BibitemShut {NoStop}%
\bibitem [{\citenamefont {Manzano}\ \emph {et~al.}(2018)\citenamefont
  {Manzano}, \citenamefont {Plastina},\ and\ \citenamefont
  {Zambrini}}]{Manzano2018}%
  \BibitemOpen
  \bibfield  {author} {\bibinfo {author} {\bibfnamefont {G.}~\bibnamefont
  {Manzano}}, \bibinfo {author} {\bibfnamefont {F.}~\bibnamefont {Plastina}},\
  and\ \bibinfo {author} {\bibfnamefont {R.}~\bibnamefont {Zambrini}},\
  }\bibfield  {title} {\bibinfo {title} {Optimal work extraction and
  thermodynamics of quantum measurements and correlations},\ }\href
  {https://doi.org/10.1103/PhysRevLett.121.120602} {\bibfield  {journal}
  {\bibinfo  {journal} {Phys. Rev. Lett.}\ }\textbf {\bibinfo {volume} {121}},\
  \bibinfo {pages} {120602} (\bibinfo {year} {2018})}\BibitemShut {NoStop}%
\bibitem [{\citenamefont {Woods}\ \emph {et~al.}(2019)\citenamefont {Woods},
  \citenamefont {Ng},\ and\ \citenamefont
  {Wehner}}]{Woods2019maximumefficiencyof}%
  \BibitemOpen
  \bibfield  {author} {\bibinfo {author} {\bibfnamefont {M.~P.}\ \bibnamefont
  {Woods}}, \bibinfo {author} {\bibfnamefont {N.~H.~Y.}\ \bibnamefont {Ng}},\
  and\ \bibinfo {author} {\bibfnamefont {S.}~\bibnamefont {Wehner}},\
  }\bibfield  {title} {\bibinfo {title} {The maximum efficiency of nano heat
  engines depends on more than temperature},\ }\href
  {https://doi.org/10.22331/q-2019-08-19-177} {\bibfield  {journal} {\bibinfo
  {journal} {{Quantum}}\ }\textbf {\bibinfo {volume} {3}},\ \bibinfo {pages}
  {177} (\bibinfo {year} {2019})}\BibitemShut {NoStop}%
\bibitem [{\citenamefont {Niedenzu}\ \emph
  {et~al.}(2019{\natexlab{a}})\citenamefont {Niedenzu}, \citenamefont {Huber},\
  and\ \citenamefont {Boukobza}}]{Niedenzu2019conceptsofworkin}%
  \BibitemOpen
  \bibfield  {author} {\bibinfo {author} {\bibfnamefont {W.}~\bibnamefont
  {Niedenzu}}, \bibinfo {author} {\bibfnamefont {M.}~\bibnamefont {Huber}},\
  and\ \bibinfo {author} {\bibfnamefont {E.}~\bibnamefont {Boukobza}},\
  }\bibfield  {title} {\bibinfo {title} {Concepts of work in autonomous quantum
  heat engines},\ }\href {https://doi.org/10.22331/q-2019-10-14-195} {\bibfield
   {journal} {\bibinfo  {journal} {{Quantum}}\ }\textbf {\bibinfo {volume}
  {3}},\ \bibinfo {pages} {195} (\bibinfo {year}
  {2019}{\natexlab{a}})}\BibitemShut {NoStop}%
\bibitem [{\citenamefont {Horne}\ \emph {et~al.}(2020)\citenamefont {Horne},
  \citenamefont {Yum}, \citenamefont {Dutta}, \citenamefont {H\"{a}nggi},
  \citenamefont {Gong}, \citenamefont {Poletti},\ and\ \citenamefont
  {Mukherjee}}]{VanHorne2020}%
  \BibitemOpen
  \bibfield  {author} {\bibinfo {author} {\bibfnamefont {N.~V.}\ \bibnamefont
  {Horne}}, \bibinfo {author} {\bibfnamefont {D.}~\bibnamefont {Yum}}, \bibinfo
  {author} {\bibfnamefont {T.}~\bibnamefont {Dutta}}, \bibinfo {author}
  {\bibfnamefont {P.}~\bibnamefont {H\"{a}nggi}}, \bibinfo {author}
  {\bibfnamefont {J.}~\bibnamefont {Gong}}, \bibinfo {author} {\bibfnamefont
  {D.}~\bibnamefont {Poletti}},\ and\ \bibinfo {author} {\bibfnamefont
  {M.}~\bibnamefont {Mukherjee}},\ }\bibfield  {title} {\bibinfo {title}
  {Single-atom energy-conversion device with a quantum load},\ }\href
  {https://doi.org/10.1038/s41534-020-0264-6} {\bibfield  {journal} {\bibinfo
  {journal} {npj Quantum Inf.}\ }\textbf {\bibinfo {volume} {6}} (\bibinfo
  {year} {2020})}\BibitemShut {NoStop}%
\bibitem [{\citenamefont {Gluza}\ \emph {et~al.}(2021)\citenamefont {Gluza},
  \citenamefont {Sabino}, \citenamefont {Ng}, \citenamefont {Vitagliano},
  \citenamefont {Pezzutto}, \citenamefont {Omar}, \citenamefont {Mazets},
  \citenamefont {Huber}, \citenamefont {Schmiedmayer},\ and\ \citenamefont
  {Eisert}}]{Gluza2021PRXQuantum}%
  \BibitemOpen
  \bibfield  {author} {\bibinfo {author} {\bibfnamefont {M.}~\bibnamefont
  {Gluza}}, \bibinfo {author} {\bibfnamefont {J.}~\bibnamefont {Sabino}},
  \bibinfo {author} {\bibfnamefont {N.~H.~Y.}\ \bibnamefont {Ng}}, \bibinfo
  {author} {\bibfnamefont {G.}~\bibnamefont {Vitagliano}}, \bibinfo {author}
  {\bibfnamefont {M.}~\bibnamefont {Pezzutto}}, \bibinfo {author}
  {\bibfnamefont {Y.}~\bibnamefont {Omar}}, \bibinfo {author} {\bibfnamefont
  {I.}~\bibnamefont {Mazets}}, \bibinfo {author} {\bibfnamefont
  {M.}~\bibnamefont {Huber}}, \bibinfo {author} {\bibfnamefont
  {J.}~\bibnamefont {Schmiedmayer}},\ and\ \bibinfo {author} {\bibfnamefont
  {J.}~\bibnamefont {Eisert}},\ }\bibfield  {title} {\bibinfo {title} {Quantum
  field thermal machines},\ }\href
  {https://doi.org/10.1103/PRXQuantum.2.030310} {\bibfield  {journal} {\bibinfo
   {journal} {PRX Quantum}\ }\textbf {\bibinfo {volume} {2}},\ \bibinfo {pages}
  {030310} (\bibinfo {year} {2021})}\BibitemShut {NoStop}%
\bibitem [{\citenamefont {Son}\ \emph {et~al.}(2021)\citenamefont {Son},
  \citenamefont {Talkner},\ and\ \citenamefont
  {Thingna}}]{Jeongrak2021PRXQuantum}%
  \BibitemOpen
  \bibfield  {author} {\bibinfo {author} {\bibfnamefont {J.}~\bibnamefont
  {Son}}, \bibinfo {author} {\bibfnamefont {P.}~\bibnamefont {Talkner}},\ and\
  \bibinfo {author} {\bibfnamefont {J.}~\bibnamefont {Thingna}},\ }\bibfield
  {title} {\bibinfo {title} {Monitoring quantum otto engines},\ }\href
  {https://doi.org/10.1103/PRXQuantum.2.040328} {\bibfield  {journal} {\bibinfo
   {journal} {PRX Quantum}\ }\textbf {\bibinfo {volume} {2}},\ \bibinfo {pages}
  {040328} (\bibinfo {year} {2021})}\BibitemShut {NoStop}%
\bibitem [{\citenamefont {Lu}\ \emph {et~al.}(2022)\citenamefont {Lu},
  \citenamefont {Lambert}, \citenamefont {Kockum}, \citenamefont {Funo},
  \citenamefont {Bengtsson}, \citenamefont {Gasparinetti}, \citenamefont
  {Nori},\ and\ \citenamefont {Delsing}}]{Lu2022PRXQuantum}%
  \BibitemOpen
  \bibfield  {author} {\bibinfo {author} {\bibfnamefont {Y.}~\bibnamefont
  {Lu}}, \bibinfo {author} {\bibfnamefont {N.}~\bibnamefont {Lambert}},
  \bibinfo {author} {\bibfnamefont {A.~F.}\ \bibnamefont {Kockum}}, \bibinfo
  {author} {\bibfnamefont {K.}~\bibnamefont {Funo}}, \bibinfo {author}
  {\bibfnamefont {A.}~\bibnamefont {Bengtsson}}, \bibinfo {author}
  {\bibfnamefont {S.}~\bibnamefont {Gasparinetti}}, \bibinfo {author}
  {\bibfnamefont {F.}~\bibnamefont {Nori}},\ and\ \bibinfo {author}
  {\bibfnamefont {P.}~\bibnamefont {Delsing}},\ }\bibfield  {title} {\bibinfo
  {title} {Steady-state heat transport and work with a single artificial atom
  coupled to a waveguide: Emission without external driving},\ }\href
  {https://doi.org/10.1103/PRXQuantum.3.020305} {\bibfield  {journal} {\bibinfo
   {journal} {PRX Quantum}\ }\textbf {\bibinfo {volume} {3}},\ \bibinfo {pages}
  {020305} (\bibinfo {year} {2022})}\BibitemShut {NoStop}%
\bibitem [{\citenamefont {Funo}\ \emph {et~al.}(2013)\citenamefont {Funo},
  \citenamefont {Watanabe},\ and\ \citenamefont {Ueda}}]{Funo2013}%
  \BibitemOpen
  \bibfield  {author} {\bibinfo {author} {\bibfnamefont {K.}~\bibnamefont
  {Funo}}, \bibinfo {author} {\bibfnamefont {Y.}~\bibnamefont {Watanabe}},\
  and\ \bibinfo {author} {\bibfnamefont {M.}~\bibnamefont {Ueda}},\ }\bibfield
  {title} {\bibinfo {title} {Thermodynamic work gain from entanglement},\
  }\href {https://doi.org/10.1103/PhysRevA.88.052319} {\bibfield  {journal}
  {\bibinfo  {journal} {Phys. Rev. A}\ }\textbf {\bibinfo {volume} {88}},\
  \bibinfo {pages} {052319} (\bibinfo {year} {2013})}\BibitemShut {NoStop}%
\bibitem [{\citenamefont {Skrzypczyk}\ \emph {et~al.}(2014)\citenamefont
  {Skrzypczyk}, \citenamefont {Short},\ and\ \citenamefont
  {Popescu}}]{Skrzypczyk2014}%
  \BibitemOpen
  \bibfield  {author} {\bibinfo {author} {\bibfnamefont {P.}~\bibnamefont
  {Skrzypczyk}}, \bibinfo {author} {\bibfnamefont {A.~J.}\ \bibnamefont
  {Short}},\ and\ \bibinfo {author} {\bibfnamefont {S.}~\bibnamefont
  {Popescu}},\ }\bibfield  {title} {\bibinfo {title} {Work extraction and
  thermodynamics for individual quantum systems},\ }\href
  {https://doi.org/10.1038/ncomms5185} {\bibfield  {journal} {\bibinfo
  {journal} {Nat. Commun.}\ }\textbf {\bibinfo {volume} {5}} (\bibinfo {year}
  {2014})}\BibitemShut {NoStop}%
\bibitem [{\citenamefont {Alhambra}\ \emph {et~al.}(2016)\citenamefont
  {Alhambra}, \citenamefont {Masanes}, \citenamefont {Oppenheim},\ and\
  \citenamefont {Perry}}]{Alhambra2016}%
  \BibitemOpen
  \bibfield  {author} {\bibinfo {author} {\bibfnamefont {A.~M.}\ \bibnamefont
  {Alhambra}}, \bibinfo {author} {\bibfnamefont {L.}~\bibnamefont {Masanes}},
  \bibinfo {author} {\bibfnamefont {J.}~\bibnamefont {Oppenheim}},\ and\
  \bibinfo {author} {\bibfnamefont {C.}~\bibnamefont {Perry}},\ }\bibfield
  {title} {\bibinfo {title} {Fluctuating work: From quantum thermodynamical
  identities to a second law equality},\ }\href
  {https://doi.org/10.1103/PhysRevX.6.041017} {\bibfield  {journal} {\bibinfo
  {journal} {Phys. Rev. X}\ }\textbf {\bibinfo {volume} {6}},\ \bibinfo {pages}
  {041017} (\bibinfo {year} {2016})}\BibitemShut {NoStop}%
\bibitem [{\citenamefont {Elouard}\ \emph {et~al.}(2017)\citenamefont
  {Elouard}, \citenamefont {Herrera-Mart\'{\i}}, \citenamefont {Huard},\ and\
  \citenamefont {Auff\`eves}}]{Elouard2017}%
  \BibitemOpen
  \bibfield  {author} {\bibinfo {author} {\bibfnamefont {C.}~\bibnamefont
  {Elouard}}, \bibinfo {author} {\bibfnamefont {D.}~\bibnamefont
  {Herrera-Mart\'{\i}}}, \bibinfo {author} {\bibfnamefont {B.}~\bibnamefont
  {Huard}},\ and\ \bibinfo {author} {\bibfnamefont {A.}~\bibnamefont
  {Auff\`eves}},\ }\bibfield  {title} {\bibinfo {title} {Extracting work from
  quantum measurement in maxwell's demon engines},\ }\href
  {https://doi.org/10.1103/PhysRevLett.118.260603} {\bibfield  {journal}
  {\bibinfo  {journal} {Phys. Rev. Lett.}\ }\textbf {\bibinfo {volume} {118}},\
  \bibinfo {pages} {260603} (\bibinfo {year} {2017})}\BibitemShut {NoStop}%
\bibitem [{\citenamefont {Morris}\ \emph {et~al.}(2019)\citenamefont {Morris},
  \citenamefont {Lami},\ and\ \citenamefont {Adesso}}]{Morris2019}%
  \BibitemOpen
  \bibfield  {author} {\bibinfo {author} {\bibfnamefont {B.}~\bibnamefont
  {Morris}}, \bibinfo {author} {\bibfnamefont {L.}~\bibnamefont {Lami}},\ and\
  \bibinfo {author} {\bibfnamefont {G.}~\bibnamefont {Adesso}},\ }\bibfield
  {title} {\bibinfo {title} {Assisted work distillation},\ }\href
  {https://doi.org/10.1103/PhysRevLett.122.130601} {\bibfield  {journal}
  {\bibinfo  {journal} {Phys. Rev. Lett.}\ }\textbf {\bibinfo {volume} {122}},\
  \bibinfo {pages} {130601} (\bibinfo {year} {2019})}\BibitemShut {NoStop}%
\bibitem [{\citenamefont {Niedenzu}\ \emph
  {et~al.}(2019{\natexlab{b}})\citenamefont {Niedenzu}, \citenamefont {Huber},\
  and\ \citenamefont {Boukobza}}]{Niedenzu2019}%
  \BibitemOpen
  \bibfield  {author} {\bibinfo {author} {\bibfnamefont {W.}~\bibnamefont
  {Niedenzu}}, \bibinfo {author} {\bibfnamefont {M.}~\bibnamefont {Huber}},\
  and\ \bibinfo {author} {\bibfnamefont {E.}~\bibnamefont {Boukobza}},\
  }\bibfield  {title} {\bibinfo {title} {Concepts of work in autonomous quantum
  heat engines},\ }\href {https://doi.org/10.22331/q-2019-10-14-195} {\bibfield
   {journal} {\bibinfo  {journal} {Quantum}\ }\textbf {\bibinfo {volume} {3}},\
  \bibinfo {pages} {195} (\bibinfo {year} {2019}{\natexlab{b}})}\BibitemShut
  {NoStop}%
\bibitem [{\citenamefont {Allahverdyan}(2014)}]{Allahverdyan2014}%
  \BibitemOpen
  \bibfield  {author} {\bibinfo {author} {\bibfnamefont {A.~E.}\ \bibnamefont
  {Allahverdyan}},\ }\bibfield  {title} {\bibinfo {title} {Nonequilibrium
  quantum fluctuations of work},\ }\href
  {https://doi.org/10.1103/PhysRevE.90.032137} {\bibfield  {journal} {\bibinfo
  {journal} {Phys. Rev. E}\ }\textbf {\bibinfo {volume} {90}},\ \bibinfo
  {pages} {032137} (\bibinfo {year} {2014})}\BibitemShut {NoStop}%
\bibitem [{\citenamefont {Talkner}\ and\ \citenamefont
  {H\"anggi}(2016)}]{Talkner2016}%
  \BibitemOpen
  \bibfield  {author} {\bibinfo {author} {\bibfnamefont {P.}~\bibnamefont
  {Talkner}}\ and\ \bibinfo {author} {\bibfnamefont {P.}~\bibnamefont
  {H\"anggi}},\ }\bibfield  {title} {\bibinfo {title} {Aspects of quantum
  work},\ }\href {https://doi.org/10.1103/PhysRevE.93.022131} {\bibfield
  {journal} {\bibinfo  {journal} {Phys. Rev. E}\ }\textbf {\bibinfo {volume}
  {93}},\ \bibinfo {pages} {022131} (\bibinfo {year} {2016})}\BibitemShut
  {NoStop}%
\bibitem [{\citenamefont {Lostaglio}\ \emph {et~al.}(2018)\citenamefont
  {Lostaglio}, \citenamefont {Alhambra},\ and\ \citenamefont
  {Perry}}]{Lostaglio2018}%
  \BibitemOpen
  \bibfield  {author} {\bibinfo {author} {\bibfnamefont {M.}~\bibnamefont
  {Lostaglio}}, \bibinfo {author} {\bibfnamefont {{\'{A}}.~M.}\ \bibnamefont
  {Alhambra}},\ and\ \bibinfo {author} {\bibfnamefont {C.}~\bibnamefont
  {Perry}},\ }\bibfield  {title} {\bibinfo {title} {Elementary thermal
  operations},\ }\href {https://doi.org/10.22331/q-2018-02-08-52} {\bibfield
  {journal} {\bibinfo  {journal} {Quantum}\ }\textbf {\bibinfo {volume} {2}},\
  \bibinfo {pages} {52} (\bibinfo {year} {2018})}\BibitemShut {NoStop}%
\bibitem [{\citenamefont {Schrödinger}(1935)}]{Schrodinger1935}%
  \BibitemOpen
  \bibfield  {author} {\bibinfo {author} {\bibfnamefont {E.}~\bibnamefont
  {Schrödinger}},\ }\bibfield  {title} {\bibinfo {title} {Discussion of
  probability relations between separated systems},\ }\href
  {https://doi.org/10.1017/S0305004100013554} {\bibfield  {journal} {\bibinfo
  {journal} {Math. Proc. Camb. Philos. Soc.}\ }\textbf {\bibinfo {volume}
  {31}},\ \bibinfo {pages} {555–563} (\bibinfo {year} {1935})}\BibitemShut
  {NoStop}%
\bibitem [{\citenamefont {Wiseman}\ \emph {et~al.}(2007)\citenamefont
  {Wiseman}, \citenamefont {Jones},\ and\ \citenamefont
  {Doherty}}]{Wiseman2007}%
  \BibitemOpen
  \bibfield  {author} {\bibinfo {author} {\bibfnamefont {H.~M.}\ \bibnamefont
  {Wiseman}}, \bibinfo {author} {\bibfnamefont {S.~J.}\ \bibnamefont {Jones}},\
  and\ \bibinfo {author} {\bibfnamefont {A.~C.}\ \bibnamefont {Doherty}},\
  }\bibfield  {title} {\bibinfo {title} {Steering, entanglement, nonlocality,
  and the einstein-podolsky-rosen paradox},\ }\href
  {https://doi.org/10.1103/PhysRevLett.98.140402} {\bibfield  {journal}
  {\bibinfo  {journal} {Phys. Rev. Lett.}\ }\textbf {\bibinfo {volume} {98}},\
  \bibinfo {pages} {140402} (\bibinfo {year} {2007})}\BibitemShut {NoStop}%
\bibitem [{\citenamefont {Cavalcanti}\ and\ \citenamefont
  {Skrzypczyk}(2016)}]{Cavalcanti2016}%
  \BibitemOpen
  \bibfield  {author} {\bibinfo {author} {\bibfnamefont {D.}~\bibnamefont
  {Cavalcanti}}\ and\ \bibinfo {author} {\bibfnamefont {P.}~\bibnamefont
  {Skrzypczyk}},\ }\bibfield  {title} {\bibinfo {title} {Quantum steering: a
  review with focus on semidefinite programming},\ }\href
  {https://doi.org/10.1088/1361-6633/80/2/024001} {\bibfield  {journal}
  {\bibinfo  {journal} {\text{Rep. Prog. Phys.}}\ }\textbf {\bibinfo {volume}
  {80}},\ \bibinfo {pages} {024001} (\bibinfo {year} {2016})}\BibitemShut
  {NoStop}%
\bibitem [{\citenamefont {Uola}\ \emph {et~al.}(2020)\citenamefont {Uola},
  \citenamefont {Costa}, \citenamefont {Nguyen},\ and\ \citenamefont
  {G\"uhne}}]{Uola2020}%
  \BibitemOpen
  \bibfield  {author} {\bibinfo {author} {\bibfnamefont {R.}~\bibnamefont
  {Uola}}, \bibinfo {author} {\bibfnamefont {A.~C.~S.}\ \bibnamefont {Costa}},
  \bibinfo {author} {\bibfnamefont {H.~C.}\ \bibnamefont {Nguyen}},\ and\
  \bibinfo {author} {\bibfnamefont {O.}~\bibnamefont {G\"uhne}},\ }\bibfield
  {title} {\bibinfo {title} {Quantum steering},\ }\href
  {https://doi.org/10.1103/RevModPhys.92.015001} {\bibfield  {journal}
  {\bibinfo  {journal} {Rev. Mod. Phys.}\ }\textbf {\bibinfo {volume} {92}},\
  \bibinfo {pages} {015001} (\bibinfo {year} {2020})}\BibitemShut {NoStop}%
\bibitem [{\citenamefont {Xiang}\ \emph {et~al.}(2022)\citenamefont {Xiang},
  \citenamefont {Cheng}, \citenamefont {Gong}, \citenamefont {Ficek},\ and\
  \citenamefont {He}}]{Xiang2022}%
  \BibitemOpen
  \bibfield  {author} {\bibinfo {author} {\bibfnamefont {Y.}~\bibnamefont
  {Xiang}}, \bibinfo {author} {\bibfnamefont {S.}~\bibnamefont {Cheng}},
  \bibinfo {author} {\bibfnamefont {Q.}~\bibnamefont {Gong}}, \bibinfo {author}
  {\bibfnamefont {Z.}~\bibnamefont {Ficek}},\ and\ \bibinfo {author}
  {\bibfnamefont {Q.}~\bibnamefont {He}},\ }\bibfield  {title} {\bibinfo
  {title} {Quantum steering: Practical challenges and future directions},\
  }\href {https://doi.org/10.1103/PRXQuantum.3.030102} {\bibfield  {journal}
  {\bibinfo  {journal} {PRX Quantum}\ }\textbf {\bibinfo {volume} {3}},\
  \bibinfo {pages} {030102} (\bibinfo {year} {2022})}\BibitemShut {NoStop}%
\bibitem [{\citenamefont {Beyer}\ \emph {et~al.}(2019)\citenamefont {Beyer},
  \citenamefont {Luoma},\ and\ \citenamefont {Strunz}}]{Beyer2019}%
  \BibitemOpen
  \bibfield  {author} {\bibinfo {author} {\bibfnamefont {K.}~\bibnamefont
  {Beyer}}, \bibinfo {author} {\bibfnamefont {K.}~\bibnamefont {Luoma}},\ and\
  \bibinfo {author} {\bibfnamefont {W.~T.}\ \bibnamefont {Strunz}},\ }\bibfield
   {title} {\bibinfo {title} {Steering heat engines: A truly quantum {M}axwell
  demon},\ }\href {https://doi.org/10.1103/PhysRevLett.123.250606} {\bibfield
  {journal} {\bibinfo  {journal} {Phys. Rev. Lett.}\ }\textbf {\bibinfo
  {volume} {123}},\ \bibinfo {pages} {250606} (\bibinfo {year}
  {2019})}\BibitemShut {NoStop}%
\bibitem [{\citenamefont {Ji}\ \emph {et~al.}(2022)\citenamefont {Ji},
  \citenamefont {Chai}, \citenamefont {Wang}, \citenamefont {Guo},
  \citenamefont {Rong}, \citenamefont {Shi}, \citenamefont {Ren}, \citenamefont
  {Wang},\ and\ \citenamefont {Du}}]{Ji2022}%
  \BibitemOpen
  \bibfield  {author} {\bibinfo {author} {\bibfnamefont {W.}~\bibnamefont
  {Ji}}, \bibinfo {author} {\bibfnamefont {Z.}~\bibnamefont {Chai}}, \bibinfo
  {author} {\bibfnamefont {M.}~\bibnamefont {Wang}}, \bibinfo {author}
  {\bibfnamefont {Y.}~\bibnamefont {Guo}}, \bibinfo {author} {\bibfnamefont
  {X.}~\bibnamefont {Rong}}, \bibinfo {author} {\bibfnamefont {F.}~\bibnamefont
  {Shi}}, \bibinfo {author} {\bibfnamefont {C.}~\bibnamefont {Ren}}, \bibinfo
  {author} {\bibfnamefont {Y.}~\bibnamefont {Wang}},\ and\ \bibinfo {author}
  {\bibfnamefont {J.}~\bibnamefont {Du}},\ }\bibfield  {title} {\bibinfo
  {title} {Spin quantum heat engine quantified by quantum steering},\ }\href
  {https://doi.org/10.1103/PhysRevLett.128.090602} {\bibfield  {journal}
  {\bibinfo  {journal} {Phys. Rev. Lett.}\ }\textbf {\bibinfo {volume} {128}},\
  \bibinfo {pages} {090602} (\bibinfo {year} {2022})}\BibitemShut {NoStop}%
\bibitem [{\citenamefont {Knott}(1911)}]{Maxwell1871}%
  \BibitemOpen
  \bibfield  {author} {\bibinfo {author} {\bibfnamefont {C.~G.}\ \bibnamefont
  {Knott}},\ }\href@noop {} {\emph {\bibinfo {title} {Life and Scientific Work
  of Peter Guthrie Tait}}}\ (\bibinfo  {publisher} {Cambridge University
  Press},\ \bibinfo {year} {1911})\BibitemShut {NoStop}%
\bibitem [{\citenamefont {Leggett}\ and\ \citenamefont
  {Garg}(1985)}]{Leggett85}%
  \BibitemOpen
  \bibfield  {author} {\bibinfo {author} {\bibfnamefont {A.~J.}\ \bibnamefont
  {Leggett}}\ and\ \bibinfo {author} {\bibfnamefont {A.}~\bibnamefont {Garg}},\
  }\bibfield  {title} {\bibinfo {title} {Quantum mechanics versus macroscopic
  realism: Is the flux there when nobody looks?},\ }\href
  {http://link.aps.org/doi/10.1103/PhysRevLett.54.857} {\bibfield  {journal}
  {\bibinfo  {journal} {Phys. Rev. Lett.}\ }\textbf {\bibinfo {volume} {54}},\
  \bibinfo {pages} {857} (\bibinfo {year} {1985})}\BibitemShut {NoStop}%
\bibitem [{\citenamefont {Joarder}\ \emph {et~al.}(2022)\citenamefont
  {Joarder}, \citenamefont {Saha}, \citenamefont {Home},\ and\ \citenamefont
  {Sinha}}]{Joarder2022PRXQuantum}%
  \BibitemOpen
  \bibfield  {author} {\bibinfo {author} {\bibfnamefont {K.}~\bibnamefont
  {Joarder}}, \bibinfo {author} {\bibfnamefont {D.}~\bibnamefont {Saha}},
  \bibinfo {author} {\bibfnamefont {D.}~\bibnamefont {Home}},\ and\ \bibinfo
  {author} {\bibfnamefont {U.}~\bibnamefont {Sinha}},\ }\bibfield  {title}
  {\bibinfo {title} {Loophole-free interferometric test of macrorealism using
  heralded single photons},\ }\href
  {https://doi.org/10.1103/PRXQuantum.3.010307} {\bibfield  {journal} {\bibinfo
   {journal} {PRX Quantum}\ }\textbf {\bibinfo {volume} {3}},\ \bibinfo {pages}
  {010307} (\bibinfo {year} {2022})}\BibitemShut {NoStop}%
\bibitem [{\citenamefont {Emary}\ \emph {et~al.}(2014)\citenamefont {Emary},
  \citenamefont {Lambert},\ and\ \citenamefont {Nori}}]{Emary14}%
  \BibitemOpen
  \bibfield  {author} {\bibinfo {author} {\bibfnamefont {C.}~\bibnamefont
  {Emary}}, \bibinfo {author} {\bibfnamefont {N.}~\bibnamefont {Lambert}},\
  and\ \bibinfo {author} {\bibfnamefont {F.}~\bibnamefont {Nori}},\ }\bibfield
  {title} {\bibinfo {title} {{L}eggett-{G}arg inequalities},\ }\href
  {http://stacks.iop.org/0034-4885/77/i=1/a=016001} {\bibfield  {journal}
  {\bibinfo  {journal} {Rep. Prog. Phys.}\ }\textbf {\bibinfo {volume} {77}},\
  \bibinfo {pages} {016001} (\bibinfo {year} {2014})}\BibitemShut {NoStop}%
\bibitem [{\citenamefont {Ku}\ \emph {et~al.}(2022{\natexlab{a}})\citenamefont
  {Ku}, \citenamefont {Kadlec}, \citenamefont {\ifmmode~\check{C}\else
  \v{C}\fi{}ernoch}, \citenamefont {Quintino}, \citenamefont {Zhou},
  \citenamefont {Lemr}, \citenamefont {Lambert}, \citenamefont {Miranowicz},
  \citenamefont {Chen}, \citenamefont {Nori},\ and\ \citenamefont
  {Chen}}]{Ku2021}%
  \BibitemOpen
  \bibfield  {author} {\bibinfo {author} {\bibfnamefont {H.-Y.}\ \bibnamefont
  {Ku}}, \bibinfo {author} {\bibfnamefont {J.}~\bibnamefont {Kadlec}}, \bibinfo
  {author} {\bibfnamefont {A.}~\bibnamefont {\ifmmode~\check{C}\else
  \v{C}\fi{}ernoch}}, \bibinfo {author} {\bibfnamefont {M.~T.}\ \bibnamefont
  {Quintino}}, \bibinfo {author} {\bibfnamefont {W.}~\bibnamefont {Zhou}},
  \bibinfo {author} {\bibfnamefont {K.}~\bibnamefont {Lemr}}, \bibinfo {author}
  {\bibfnamefont {N.}~\bibnamefont {Lambert}}, \bibinfo {author} {\bibfnamefont
  {A.}~\bibnamefont {Miranowicz}}, \bibinfo {author} {\bibfnamefont {S.-L.}\
  \bibnamefont {Chen}}, \bibinfo {author} {\bibfnamefont {F.}~\bibnamefont
  {Nori}},\ and\ \bibinfo {author} {\bibfnamefont {Y.-N.}\ \bibnamefont
  {Chen}},\ }\bibfield  {title} {\bibinfo {title} {Quantifying quantumness of
  channels without entanglement},\ }\href
  {https://doi.org/10.1103/PRXQuantum.3.020338} {\bibfield  {journal} {\bibinfo
   {journal} {PRX Quantum}\ }\textbf {\bibinfo {volume} {3}},\ \bibinfo {pages}
  {020338} (\bibinfo {year} {2022}{\natexlab{a}})}\BibitemShut {NoStop}%
\bibitem [{\citenamefont {Vieira}\ and\ \citenamefont
  {Budroni}(2022)}]{Vieira2022temporal}%
  \BibitemOpen
  \bibfield  {author} {\bibinfo {author} {\bibfnamefont {L.~B.}\ \bibnamefont
  {Vieira}}\ and\ \bibinfo {author} {\bibfnamefont {C.}~\bibnamefont
  {Budroni}},\ }\bibfield  {title} {\bibinfo {title} {Temporal correlations in
  the simplest measurement sequences},\ }\href
  {https://doi.org/10.22331/q-2022-01-18-623} {\bibfield  {journal} {\bibinfo
  {journal} {{Quantum}}\ }\textbf {\bibinfo {volume} {6}},\ \bibinfo {pages}
  {623} (\bibinfo {year} {2022})}\BibitemShut {NoStop}%
\bibitem [{\citenamefont {Maity}\ \emph {et~al.}(2021)\citenamefont {Maity},
  \citenamefont {Mal}, \citenamefont {Jebarathinam},\ and\ \citenamefont
  {Majumdar}}]{Maity2021}%
  \BibitemOpen
  \bibfield  {author} {\bibinfo {author} {\bibfnamefont {A.~G.}\ \bibnamefont
  {Maity}}, \bibinfo {author} {\bibfnamefont {S.}~\bibnamefont {Mal}}, \bibinfo
  {author} {\bibfnamefont {C.}~\bibnamefont {Jebarathinam}},\ and\ \bibinfo
  {author} {\bibfnamefont {A.~S.}\ \bibnamefont {Majumdar}},\ }\bibfield
  {title} {\bibinfo {title} {Self-testing of binary pauli measurements
  requiring neither entanglement nor any dimensional restriction},\ }\href
  {https://doi.org/10.1103/PhysRevA.103.062604} {\bibfield  {journal} {\bibinfo
   {journal} {Phys. Rev. A}\ }\textbf {\bibinfo {volume} {103}},\ \bibinfo
  {pages} {062604} (\bibinfo {year} {2021})}\BibitemShut {NoStop}%
\bibitem [{\citenamefont {Chen}\ \emph {et~al.}(2014)\citenamefont {Chen},
  \citenamefont {Li}, \citenamefont {Lambert}, \citenamefont {Chen},
  \citenamefont {Ota}, \citenamefont {Chen},\ and\ \citenamefont
  {Nori}}]{Chen2014}%
  \BibitemOpen
  \bibfield  {author} {\bibinfo {author} {\bibfnamefont {Y.-N.}\ \bibnamefont
  {Chen}}, \bibinfo {author} {\bibfnamefont {C.-M.}\ \bibnamefont {Li}},
  \bibinfo {author} {\bibfnamefont {N.}~\bibnamefont {Lambert}}, \bibinfo
  {author} {\bibfnamefont {S.-L.}\ \bibnamefont {Chen}}, \bibinfo {author}
  {\bibfnamefont {Y.}~\bibnamefont {Ota}}, \bibinfo {author} {\bibfnamefont
  {G.-Y.}\ \bibnamefont {Chen}},\ and\ \bibinfo {author} {\bibfnamefont
  {F.}~\bibnamefont {Nori}},\ }\bibfield  {title} {\bibinfo {title} {Temporal
  steering inequality},\ }\href {https://doi.org/10.1103/PhysRevA.89.032112}
  {\bibfield  {journal} {\bibinfo  {journal} {Phys. Rev. A}\ }\textbf {\bibinfo
  {volume} {89}},\ \bibinfo {pages} {032112} (\bibinfo {year}
  {2014})}\BibitemShut {NoStop}%
\bibitem [{\citenamefont {Li}\ \emph {et~al.}(2015)\citenamefont {Li},
  \citenamefont {Chen}, \citenamefont {Lambert}, \citenamefont {Chiu},\ and\
  \citenamefont {Nori}}]{Li2015}%
  \BibitemOpen
  \bibfield  {author} {\bibinfo {author} {\bibfnamefont {C.-M.}\ \bibnamefont
  {Li}}, \bibinfo {author} {\bibfnamefont {Y.-N.}\ \bibnamefont {Chen}},
  \bibinfo {author} {\bibfnamefont {N.}~\bibnamefont {Lambert}}, \bibinfo
  {author} {\bibfnamefont {C.-Y.}\ \bibnamefont {Chiu}},\ and\ \bibinfo
  {author} {\bibfnamefont {F.}~\bibnamefont {Nori}},\ }\bibfield  {title}
  {\bibinfo {title} {Certifying single-system steering for quantum-information
  processing},\ }\href {https://doi.org/10.1103/PhysRevA.92.062310} {\bibfield
  {journal} {\bibinfo  {journal} {Phys. Rev. A}\ }\textbf {\bibinfo {volume}
  {92}},\ \bibinfo {pages} {062310} (\bibinfo {year} {2015})}\BibitemShut
  {NoStop}%
\bibitem [{\citenamefont {Chen}\ \emph
  {et~al.}(2016{\natexlab{b}})\citenamefont {Chen}, \citenamefont {Lambert},
  \citenamefont {Li}, \citenamefont {Miranowicz}, \citenamefont {Chen},\ and\
  \citenamefont {Nori}}]{Chen2016}%
  \BibitemOpen
  \bibfield  {author} {\bibinfo {author} {\bibfnamefont {S.-L.}\ \bibnamefont
  {Chen}}, \bibinfo {author} {\bibfnamefont {N.}~\bibnamefont {Lambert}},
  \bibinfo {author} {\bibfnamefont {C.-M.}\ \bibnamefont {Li}}, \bibinfo
  {author} {\bibfnamefont {A.}~\bibnamefont {Miranowicz}}, \bibinfo {author}
  {\bibfnamefont {Y.-N.}\ \bibnamefont {Chen}},\ and\ \bibinfo {author}
  {\bibfnamefont {F.}~\bibnamefont {Nori}},\ }\bibfield  {title} {\bibinfo
  {title} {Quantifying non-markovianity with temporal steering},\ }\href
  {https://doi.org/10.1103/PhysRevLett.116.020503} {\bibfield  {journal}
  {\bibinfo  {journal} {Phys. Rev. Lett.}\ }\textbf {\bibinfo {volume} {116}},\
  \bibinfo {pages} {020503} (\bibinfo {year} {2016}{\natexlab{b}})}\BibitemShut
  {NoStop}%
\bibitem [{\citenamefont {Ku}\ \emph {et~al.}(2018)\citenamefont {Ku},
  \citenamefont {Chen}, \citenamefont {Lambert}, \citenamefont {Chen},\ and\
  \citenamefont {Nori}}]{Ku2018}%
  \BibitemOpen
  \bibfield  {author} {\bibinfo {author} {\bibfnamefont {H.-Y.}\ \bibnamefont
  {Ku}}, \bibinfo {author} {\bibfnamefont {S.-L.}\ \bibnamefont {Chen}},
  \bibinfo {author} {\bibfnamefont {N.}~\bibnamefont {Lambert}}, \bibinfo
  {author} {\bibfnamefont {Y.-N.}\ \bibnamefont {Chen}},\ and\ \bibinfo
  {author} {\bibfnamefont {F.}~\bibnamefont {Nori}},\ }\bibfield  {title}
  {\bibinfo {title} {Hierarchy in temporal quantum correlations},\ }\href
  {https://doi.org/10.1103/PhysRevA.98.022104} {\bibfield  {journal} {\bibinfo
  {journal} {Phys. Rev. A}\ }\textbf {\bibinfo {volume} {98}},\ \bibinfo
  {pages} {022104} (\bibinfo {year} {2018})}\BibitemShut {NoStop}%
\bibitem [{\citenamefont {Lin}\ \emph {et~al.}(2021)\citenamefont {Lin},
  \citenamefont {Lin}, \citenamefont {Ku}, \citenamefont {Lambert},
  \citenamefont {Chen},\ and\ \citenamefont {Nori}}]{Lin2021}%
  \BibitemOpen
  \bibfield  {author} {\bibinfo {author} {\bibfnamefont {J.-D.}\ \bibnamefont
  {Lin}}, \bibinfo {author} {\bibfnamefont {W.-Y.}\ \bibnamefont {Lin}},
  \bibinfo {author} {\bibfnamefont {H.-Y.}\ \bibnamefont {Ku}}, \bibinfo
  {author} {\bibfnamefont {N.}~\bibnamefont {Lambert}}, \bibinfo {author}
  {\bibfnamefont {Y.-N.}\ \bibnamefont {Chen}},\ and\ \bibinfo {author}
  {\bibfnamefont {F.}~\bibnamefont {Nori}},\ }\bibfield  {title} {\bibinfo
  {title} {Quantum steering as a witness of quantum scrambling},\ }\href
  {https://doi.org/10.1103/PhysRevA.104.022614} {\bibfield  {journal} {\bibinfo
   {journal} {Phys. Rev. A}\ }\textbf {\bibinfo {volume} {104}},\ \bibinfo
  {pages} {022614} (\bibinfo {year} {2021})}\BibitemShut {NoStop}%
\bibitem [{\citenamefont {Gu{\'{e}}rin}\ \emph {et~al.}(2016)\citenamefont
  {Gu{\'{e}}rin}, \citenamefont {Feix}, \citenamefont {Ara{\'{u}}jo},\ and\
  \citenamefont {Brukner}}]{Guerin2016}%
  \BibitemOpen
  \bibfield  {author} {\bibinfo {author} {\bibfnamefont {P.~A.}\ \bibnamefont
  {Gu{\'{e}}rin}}, \bibinfo {author} {\bibfnamefont {A.}~\bibnamefont {Feix}},
  \bibinfo {author} {\bibfnamefont {M.}~\bibnamefont {Ara{\'{u}}jo}},\ and\
  \bibinfo {author} {\bibfnamefont {{\v{C}}.}~\bibnamefont {Brukner}},\
  }\bibfield  {title} {\bibinfo {title} {Exponential communication complexity
  advantage from quantum superposition of the direction of communication},\
  }\href {https://doi.org/10.1103/physrevlett.117.100502} {\bibfield  {journal}
  {\bibinfo  {journal} {Phys. Rev. Lett.}\ }\textbf {\bibinfo {volume} {117}},\
  \bibinfo {pages} {100502} (\bibinfo {year} {2016})}\BibitemShut {NoStop}%
\bibitem [{\citenamefont {Chiribella}\ and\ \citenamefont
  {Kristj{\'a}nsson}(2019)}]{Chiribella2019quantum}%
  \BibitemOpen
  \bibfield  {author} {\bibinfo {author} {\bibfnamefont {G.}~\bibnamefont
  {Chiribella}}\ and\ \bibinfo {author} {\bibfnamefont {H.}~\bibnamefont
  {Kristj{\'a}nsson}},\ }\bibfield  {title} {\bibinfo {title} {Quantum shannon
  theory with superpositions of trajectories},\ }\href
  {https://royalsocietypublishing.org/doi/full/10.1098/rspa.2018.0903}
  {\bibfield  {journal} {\bibinfo  {journal} {Proc. R. Soc. A}\ }\textbf
  {\bibinfo {volume} {475}},\ \bibinfo {pages} {20180903} (\bibinfo {year}
  {2019})}\BibitemShut {NoStop}%
\bibitem [{\citenamefont {Abbott}\ \emph {et~al.}(2020)\citenamefont {Abbott},
  \citenamefont {Wechs}, \citenamefont {Horsman}, \citenamefont {Mhalla},\ and\
  \citenamefont {Branciard}}]{Abbott2020}%
  \BibitemOpen
  \bibfield  {author} {\bibinfo {author} {\bibfnamefont {A.~A.}\ \bibnamefont
  {Abbott}}, \bibinfo {author} {\bibfnamefont {J.}~\bibnamefont {Wechs}},
  \bibinfo {author} {\bibfnamefont {D.}~\bibnamefont {Horsman}}, \bibinfo
  {author} {\bibfnamefont {M.}~\bibnamefont {Mhalla}},\ and\ \bibinfo {author}
  {\bibfnamefont {C.}~\bibnamefont {Branciard}},\ }\bibfield  {title} {\bibinfo
  {title} {Communication through coherent control of quantum channels},\ }\href
  {https://doi.org/10.22331/q-2020-09-24-333} {\bibfield  {journal} {\bibinfo
  {journal} {Quantum}\ }\textbf {\bibinfo {volume} {4}},\ \bibinfo {pages}
  {333} (\bibinfo {year} {2020})}\BibitemShut {NoStop}%
\bibitem [{\citenamefont {Kristj{\'a}nsson}\ \emph {et~al.}(2020)\citenamefont
  {Kristj{\'a}nsson}, \citenamefont {Chiribella}, \citenamefont {Salek},
  \citenamefont {Ebler},\ and\ \citenamefont
  {Wilson}}]{kristjansson2020resource}%
  \BibitemOpen
  \bibfield  {author} {\bibinfo {author} {\bibfnamefont {H.}~\bibnamefont
  {Kristj{\'a}nsson}}, \bibinfo {author} {\bibfnamefont {G.}~\bibnamefont
  {Chiribella}}, \bibinfo {author} {\bibfnamefont {S.}~\bibnamefont {Salek}},
  \bibinfo {author} {\bibfnamefont {D.}~\bibnamefont {Ebler}},\ and\ \bibinfo
  {author} {\bibfnamefont {M.}~\bibnamefont {Wilson}},\ }\bibfield  {title}
  {\bibinfo {title} {Resource theories of communication},\ }\href
  {https://iopscience.iop.org/article/10.1088/1367-2630/ab8ef7/meta} {\bibfield
   {journal} {\bibinfo  {journal} {New J. Phys.}\ }\textbf {\bibinfo {volume}
  {22}},\ \bibinfo {pages} {073014} (\bibinfo {year} {2020})}\BibitemShut
  {NoStop}%
\bibitem [{\citenamefont {Ban}(2021)}]{Ban2021two}%
  \BibitemOpen
  \bibfield  {author} {\bibinfo {author} {\bibfnamefont {M.}~\bibnamefont
  {Ban}},\ }\bibfield  {title} {\bibinfo {title} {Two-qubit correlation in two
  independent environments with indefiniteness},\ }\href
  {https://www.sciencedirect.com/science/article/pii/S0375960120308033}
  {\bibfield  {journal} {\bibinfo  {journal} {Phys. Lett. A.}\ }\textbf
  {\bibinfo {volume} {385}},\ \bibinfo {pages} {126936} (\bibinfo {year}
  {2021})}\BibitemShut {NoStop}%
\bibitem [{\citenamefont {Rubino}\ \emph {et~al.}(2021)\citenamefont {Rubino},
  \citenamefont {Rozema}, \citenamefont {Ebler}, \citenamefont
  {Kristj\'ansson}, \citenamefont {Salek}, \citenamefont {Allard~Gu\'erin},
  \citenamefont {Abbott}, \citenamefont {Branciard}, \citenamefont {Brukner},
  \citenamefont {Chiribella},\ and\ \citenamefont {Walther}}]{Rubino2021}%
  \BibitemOpen
  \bibfield  {author} {\bibinfo {author} {\bibfnamefont {G.}~\bibnamefont
  {Rubino}}, \bibinfo {author} {\bibfnamefont {L.~A.}\ \bibnamefont {Rozema}},
  \bibinfo {author} {\bibfnamefont {D.}~\bibnamefont {Ebler}}, \bibinfo
  {author} {\bibfnamefont {H.}~\bibnamefont {Kristj\'ansson}}, \bibinfo
  {author} {\bibfnamefont {S.}~\bibnamefont {Salek}}, \bibinfo {author}
  {\bibfnamefont {P.}~\bibnamefont {Allard~Gu\'erin}}, \bibinfo {author}
  {\bibfnamefont {A.~A.}\ \bibnamefont {Abbott}}, \bibinfo {author}
  {\bibfnamefont {C.}~\bibnamefont {Branciard}}, \bibinfo {author}
  {\bibfnamefont {{\v{C}}.}~\bibnamefont {Brukner}}, \bibinfo {author}
  {\bibfnamefont {G.}~\bibnamefont {Chiribella}},\ and\ \bibinfo {author}
  {\bibfnamefont {P.}~\bibnamefont {Walther}},\ }\bibfield  {title} {\bibinfo
  {title} {Experimental quantum communication enhancement by superposing
  trajectories},\ }\href {https://doi.org/10.1103/PhysRevResearch.3.013093}
  {\bibfield  {journal} {\bibinfo  {journal} {Phys. Rev. Research}\ }\textbf
  {\bibinfo {volume} {3}},\ \bibinfo {pages} {013093} (\bibinfo {year}
  {2021})}\BibitemShut {NoStop}%
\bibitem [{\citenamefont {Lin}\ \emph {et~al.}(2022)\citenamefont {Lin},
  \citenamefont {Huang}, \citenamefont {Lambert}, \citenamefont {Chen},
  \citenamefont {Nori},\ and\ \citenamefont {Chen}}]{Lin2022}%
  \BibitemOpen
  \bibfield  {author} {\bibinfo {author} {\bibfnamefont {J.-D.}\ \bibnamefont
  {Lin}}, \bibinfo {author} {\bibfnamefont {C.-Y.}\ \bibnamefont {Huang}},
  \bibinfo {author} {\bibfnamefont {N.}~\bibnamefont {Lambert}}, \bibinfo
  {author} {\bibfnamefont {G.-Y.}\ \bibnamefont {Chen}}, \bibinfo {author}
  {\bibfnamefont {F.}~\bibnamefont {Nori}},\ and\ \bibinfo {author}
  {\bibfnamefont {Y.-N.}\ \bibnamefont {Chen}},\ }\bibfield  {title} {\bibinfo
  {title} {Space-time dual quantum zeno effect: Interferometric engineering of
  open quantum system dynamics},\ }\href
  {https://doi.org/10.1103/PhysRevResearch.4.033143} {\bibfield  {journal}
  {\bibinfo  {journal} {Phys. Rev. Research}\ }\textbf {\bibinfo {volume}
  {4}},\ \bibinfo {pages} {033143} (\bibinfo {year} {2022})}\BibitemShut
  {NoStop}%
\bibitem [{\citenamefont {Lee}\ \emph {et~al.}(2022)\citenamefont {Lee},
  \citenamefont {Lin}, \citenamefont {Miranowicz}, \citenamefont {Nori},
  \citenamefont {Ku},\ and\ \citenamefont {Chen}}]{Lee2022}%
  \BibitemOpen
  \bibfield  {author} {\bibinfo {author} {\bibfnamefont {K.-Y.}\ \bibnamefont
  {Lee}}, \bibinfo {author} {\bibfnamefont {J.-D.}\ \bibnamefont {Lin}},
  \bibinfo {author} {\bibfnamefont {A.}~\bibnamefont {Miranowicz}}, \bibinfo
  {author} {\bibfnamefont {F.}~\bibnamefont {Nori}}, \bibinfo {author}
  {\bibfnamefont {H.-Y.}\ \bibnamefont {Ku}},\ and\ \bibinfo {author}
  {\bibfnamefont {Y.-N.}\ \bibnamefont {Chen}},\ }\bibfield  {title} {\bibinfo
  {title} {Steering-enhanced quantum metrology using superpositions of quantum
  channels},\ }\href {https://arxiv.org/abs/2206.03760} {\bibfield  {journal}
  {\bibinfo  {journal} {arXiv:2206.03760}\ } (\bibinfo {year}
  {2022})}\BibitemShut {NoStop}%
\bibitem [{IBM()}]{IBMWeb}%
  \BibitemOpen
  \href@noop {} {\bibinfo {title} {{IBM} {Q}uantum {S}ervices}},\ \bibinfo
  {howpublished}
  {\url{https://quantum-computing.ibm.com/services?services=systems&system}},\
  \bibinfo {note} {[May. 2022]}\BibitemShut {NoStop}%
\bibitem [{\citenamefont {Berke}\ \emph {et~al.}(2022)\citenamefont {Berke},
  \citenamefont {Varvelis}, \citenamefont {Trebst}, \citenamefont {Altland},\
  and\ \citenamefont {DiVincenzo}}]{Berke2022}%
  \BibitemOpen
  \bibfield  {author} {\bibinfo {author} {\bibfnamefont {C.}~\bibnamefont
  {Berke}}, \bibinfo {author} {\bibfnamefont {E.}~\bibnamefont {Varvelis}},
  \bibinfo {author} {\bibfnamefont {S.}~\bibnamefont {Trebst}}, \bibinfo
  {author} {\bibfnamefont {A.}~\bibnamefont {Altland}},\ and\ \bibinfo {author}
  {\bibfnamefont {D.~P.}\ \bibnamefont {DiVincenzo}},\ }\bibfield  {title}
  {\bibinfo {title} {Transmon platform for quantum computing challenged by
  chaotic fluctuations},\ }\href {https://doi.org/10.1038/s41467-022-29940-y}
  {\bibfield  {journal} {\bibinfo  {journal} {Nat. Commun.}\ }\textbf {\bibinfo
  {volume} {13}} (\bibinfo {year} {2022})}\BibitemShut {NoStop}%
\bibitem [{\citenamefont {Strikis}\ \emph {et~al.}(2021)\citenamefont
  {Strikis}, \citenamefont {Qin}, \citenamefont {Chen}, \citenamefont
  {Benjamin},\ and\ \citenamefont {Li}}]{Strikis2021PRXQuantum}%
  \BibitemOpen
  \bibfield  {author} {\bibinfo {author} {\bibfnamefont {A.}~\bibnamefont
  {Strikis}}, \bibinfo {author} {\bibfnamefont {D.}~\bibnamefont {Qin}},
  \bibinfo {author} {\bibfnamefont {Y.}~\bibnamefont {Chen}}, \bibinfo {author}
  {\bibfnamefont {S.~C.}\ \bibnamefont {Benjamin}},\ and\ \bibinfo {author}
  {\bibfnamefont {Y.}~\bibnamefont {Li}},\ }\bibfield  {title} {\bibinfo
  {title} {Learning-based quantum error mitigation},\ }\href
  {https://doi.org/10.1103/PRXQuantum.2.040330} {\bibfield  {journal} {\bibinfo
   {journal} {PRX Quantum}\ }\textbf {\bibinfo {volume} {2}},\ \bibinfo {pages}
  {040330} (\bibinfo {year} {2021})}\BibitemShut {NoStop}%
\bibitem [{\citenamefont {Pogorelov}\ \emph {et~al.}(2021)\citenamefont
  {Pogorelov}, \citenamefont {Feldker}, \citenamefont {Marciniak},
  \citenamefont {Postler}, \citenamefont {Jacob}, \citenamefont
  {Krieglsteiner}, \citenamefont {Podlesnic}, \citenamefont {Meth},
  \citenamefont {Negnevitsky}, \citenamefont {Stadler}, \citenamefont
  {H\"ofer}, \citenamefont {W\"achter}, \citenamefont {Lakhmanskiy},
  \citenamefont {Blatt}, \citenamefont {Schindler},\ and\ \citenamefont
  {Monz}}]{Pogorelov2021PRXQuantum}%
  \BibitemOpen
  \bibfield  {author} {\bibinfo {author} {\bibfnamefont {I.}~\bibnamefont
  {Pogorelov}}, \bibinfo {author} {\bibfnamefont {T.}~\bibnamefont {Feldker}},
  \bibinfo {author} {\bibfnamefont {C.~D.}\ \bibnamefont {Marciniak}}, \bibinfo
  {author} {\bibfnamefont {L.}~\bibnamefont {Postler}}, \bibinfo {author}
  {\bibfnamefont {G.}~\bibnamefont {Jacob}}, \bibinfo {author} {\bibfnamefont
  {O.}~\bibnamefont {Krieglsteiner}}, \bibinfo {author} {\bibfnamefont
  {V.}~\bibnamefont {Podlesnic}}, \bibinfo {author} {\bibfnamefont
  {M.}~\bibnamefont {Meth}}, \bibinfo {author} {\bibfnamefont {V.}~\bibnamefont
  {Negnevitsky}}, \bibinfo {author} {\bibfnamefont {M.}~\bibnamefont
  {Stadler}}, \bibinfo {author} {\bibfnamefont {B.}~\bibnamefont {H\"ofer}},
  \bibinfo {author} {\bibfnamefont {C.}~\bibnamefont {W\"achter}}, \bibinfo
  {author} {\bibfnamefont {K.}~\bibnamefont {Lakhmanskiy}}, \bibinfo {author}
  {\bibfnamefont {R.}~\bibnamefont {Blatt}}, \bibinfo {author} {\bibfnamefont
  {P.}~\bibnamefont {Schindler}},\ and\ \bibinfo {author} {\bibfnamefont
  {T.}~\bibnamefont {Monz}},\ }\bibfield  {title} {\bibinfo {title} {Compact
  ion-trap quantum computing demonstrator},\ }\href
  {https://doi.org/10.1103/PRXQuantum.2.020343} {\bibfield  {journal} {\bibinfo
   {journal} {PRX Quantum}\ }\textbf {\bibinfo {volume} {2}},\ \bibinfo {pages}
  {020343} (\bibinfo {year} {2021})}\BibitemShut {NoStop}%
\bibitem [{\citenamefont {Branciard}\ \emph {et~al.}(2012)\citenamefont
  {Branciard}, \citenamefont {Cavalcanti}, \citenamefont {Walborn},
  \citenamefont {Scarani},\ and\ \citenamefont {Wiseman}}]{Branciard2012}%
  \BibitemOpen
  \bibfield  {author} {\bibinfo {author} {\bibfnamefont {C.}~\bibnamefont
  {Branciard}}, \bibinfo {author} {\bibfnamefont {E.~G.}\ \bibnamefont
  {Cavalcanti}}, \bibinfo {author} {\bibfnamefont {S.~P.}\ \bibnamefont
  {Walborn}}, \bibinfo {author} {\bibfnamefont {V.}~\bibnamefont {Scarani}},\
  and\ \bibinfo {author} {\bibfnamefont {H.~M.}\ \bibnamefont {Wiseman}},\
  }\bibfield  {title} {\bibinfo {title} {One-sided device-independent quantum
  key distribution: Security, feasibility, and the connection with steering},\
  }\href {https://doi.org/10.1103/PhysRevA.85.010301} {\bibfield  {journal}
  {\bibinfo  {journal} {Phys. Rev. A}\ }\textbf {\bibinfo {volume} {85}},\
  \bibinfo {pages} {010301(R)} (\bibinfo {year} {2012})}\BibitemShut {NoStop}%
\bibitem [{\citenamefont {Yadin}\ \emph {et~al.}(2021)\citenamefont {Yadin},
  \citenamefont {Fadel},\ and\ \citenamefont {Gessner}}]{NC2021:BYadin}%
  \BibitemOpen
  \bibfield  {author} {\bibinfo {author} {\bibfnamefont {B.}~\bibnamefont
  {Yadin}}, \bibinfo {author} {\bibfnamefont {M.}~\bibnamefont {Fadel}},\ and\
  \bibinfo {author} {\bibfnamefont {M.}~\bibnamefont {Gessner}},\ }\bibfield
  {title} {\bibinfo {title} {Metrological complementarity reveals the
  {E}instein-{P}odolsky-{R}osen paradox},\ }\href
  {https://doi.org/10.1038/s41467-021-22353-3} {\bibfield  {journal} {\bibinfo
  {journal} {Nat. Commun.}\ }\textbf {\bibinfo {volume} {12}},\ \bibinfo
  {pages} {2410} (\bibinfo {year} {2021})}\BibitemShut {NoStop}%
\bibitem [{\citenamefont {Zhao}\ \emph {et~al.}(2020)\citenamefont {Zhao},
  \citenamefont {Ku}, \citenamefont {Chen}, \citenamefont {Chen}, \citenamefont
  {Nori}, \citenamefont {Xiang}, \citenamefont {Li}, \citenamefont {Guo},\ and\
  \citenamefont {Chen}}]{Zhao2020}%
  \BibitemOpen
  \bibfield  {author} {\bibinfo {author} {\bibfnamefont {Y.-Y.}\ \bibnamefont
  {Zhao}}, \bibinfo {author} {\bibfnamefont {H.-Y.}\ \bibnamefont {Ku}},
  \bibinfo {author} {\bibfnamefont {S.-L.}\ \bibnamefont {Chen}}, \bibinfo
  {author} {\bibfnamefont {H.-B.}\ \bibnamefont {Chen}}, \bibinfo {author}
  {\bibfnamefont {F.}~\bibnamefont {Nori}}, \bibinfo {author} {\bibfnamefont
  {G.-Y.}\ \bibnamefont {Xiang}}, \bibinfo {author} {\bibfnamefont {C.-F.}\
  \bibnamefont {Li}}, \bibinfo {author} {\bibfnamefont {G.-C.}\ \bibnamefont
  {Guo}},\ and\ \bibinfo {author} {\bibfnamefont {Y.-N.}\ \bibnamefont
  {Chen}},\ }\bibfield  {title} {\bibinfo {title} {Experimental demonstration
  of measurement-device-independent measure of quantum steering},\ }\href
  {https://doi.org/10.1038/s41534-020-00307-9} {\bibfield  {journal} {\bibinfo
  {journal} {npj Quantum Inf.}\ }\textbf {\bibinfo {volume} {6}},\ \bibinfo
  {pages} {77} (\bibinfo {year} {2020})}\BibitemShut {NoStop}%
\bibitem [{\citenamefont {Slussarenko}\ \emph {et~al.}(2022)\citenamefont
  {Slussarenko}, \citenamefont {Joch}, \citenamefont {Tischler}, \citenamefont
  {Ghafari}, \citenamefont {Shalm}, \citenamefont {Verma}, \citenamefont
  {Nam},\ and\ \citenamefont {Pryde}}]{Slussarenko2022}%
  \BibitemOpen
  \bibfield  {author} {\bibinfo {author} {\bibfnamefont {S.}~\bibnamefont
  {Slussarenko}}, \bibinfo {author} {\bibfnamefont {D.~J.}\ \bibnamefont
  {Joch}}, \bibinfo {author} {\bibfnamefont {N.}~\bibnamefont {Tischler}},
  \bibinfo {author} {\bibfnamefont {F.}~\bibnamefont {Ghafari}}, \bibinfo
  {author} {\bibfnamefont {L.~K.}\ \bibnamefont {Shalm}}, \bibinfo {author}
  {\bibfnamefont {V.~B.}\ \bibnamefont {Verma}}, \bibinfo {author}
  {\bibfnamefont {S.~W.}\ \bibnamefont {Nam}},\ and\ \bibinfo {author}
  {\bibfnamefont {G.~J.}\ \bibnamefont {Pryde}},\ }\bibfield  {title} {\bibinfo
  {title} {Quantum steering with vector vortex photon states with the detection
  loophole closed},\ }\href {https://doi.org/10.1038/s41534-022-00531-5}
  {\bibfield  {journal} {\bibinfo  {journal} {npj Quantum Inf.}\ }\textbf
  {\bibinfo {volume} {8}} (\bibinfo {year} {2022})}\BibitemShut {NoStop}%
\bibitem [{\citenamefont {Janzing}(2000)}]{Janzing2000}%
  \BibitemOpen
  \bibfield  {author} {\bibinfo {author} {\bibfnamefont {D.}~\bibnamefont
  {Janzing}},\ }\bibfield  {title} {\bibinfo {title} {Thermodynamic cost of
  reliability and low temperatures: Tightening landauer's principle and the
  second law},\ }\href {https://doi.org/10.1023/a:1026422630734} {\bibfield
  {journal} {\bibinfo  {journal} {Int. J. Theor. Phys.}\ }\textbf {\bibinfo
  {volume} {39}},\ \bibinfo {pages} {2717} (\bibinfo {year}
  {2000})}\BibitemShut {NoStop}%
\bibitem [{\citenamefont {Faist}\ \emph
  {et~al.}(2015{\natexlab{a}})\citenamefont {Faist}, \citenamefont
  {Oppenheim},\ and\ \citenamefont {Renner}}]{Faist2015_1}%
  \BibitemOpen
  \bibfield  {author} {\bibinfo {author} {\bibfnamefont {P.}~\bibnamefont
  {Faist}}, \bibinfo {author} {\bibfnamefont {J.}~\bibnamefont {Oppenheim}},\
  and\ \bibinfo {author} {\bibfnamefont {R.}~\bibnamefont {Renner}},\
  }\bibfield  {title} {\bibinfo {title} {Gibbs-preserving maps outperform
  thermal operations in the quantum regime},\ }\href
  {https://doi.org/10.1088/1367-2630/17/4/043003} {\bibfield  {journal}
  {\bibinfo  {journal} {New J. Phys.}\ }\textbf {\bibinfo {volume} {17}},\
  \bibinfo {pages} {043003} (\bibinfo {year} {2015}{\natexlab{a}})}\BibitemShut
  {NoStop}%
\bibitem [{\citenamefont {Faist}\ \emph
  {et~al.}(2015{\natexlab{b}})\citenamefont {Faist}, \citenamefont {Dupuis},
  \citenamefont {Oppenheim},\ and\ \citenamefont {Renner}}]{Faist2015_2}%
  \BibitemOpen
  \bibfield  {author} {\bibinfo {author} {\bibfnamefont {P.}~\bibnamefont
  {Faist}}, \bibinfo {author} {\bibfnamefont {F.}~\bibnamefont {Dupuis}},
  \bibinfo {author} {\bibfnamefont {J.}~\bibnamefont {Oppenheim}},\ and\
  \bibinfo {author} {\bibfnamefont {R.}~\bibnamefont {Renner}},\ }\bibfield
  {title} {\bibinfo {title} {The minimal work cost of information processing},\
  }\href {https://doi.org/10.1038/ncomms8669} {\bibfield  {journal} {\bibinfo
  {journal} {Nat. Commun.}\ }\textbf {\bibinfo {volume} {6}},\ \bibinfo {pages}
  {7669} (\bibinfo {year} {2015}{\natexlab{b}})}\BibitemShut {NoStop}%
\bibitem [{\citenamefont {Lostaglio}(2019)}]{Lostaglio2019}%
  \BibitemOpen
  \bibfield  {author} {\bibinfo {author} {\bibfnamefont {M.}~\bibnamefont
  {Lostaglio}},\ }\bibfield  {title} {\bibinfo {title} {An introductory review
  of the resource theory approach to thermodynamics},\ }\href
  {https://doi.org/10.1088/1361-6633/ab46e5} {\bibfield  {journal} {\bibinfo
  {journal} {Rep. Prog. Phys.}\ }\textbf {\bibinfo {volume} {82}},\ \bibinfo
  {pages} {114001} (\bibinfo {year} {2019})}\BibitemShut {NoStop}%
\bibitem [{\citenamefont {Hsieh}(2021)}]{Hsieh2021}%
  \BibitemOpen
  \bibfield  {author} {\bibinfo {author} {\bibfnamefont {C.-Y.}\ \bibnamefont
  {Hsieh}},\ }\bibfield  {title} {\bibinfo {title} {Communication, dynamical
  resource theory, and thermodynamics},\ }\href
  {https://doi.org/10.1103/PRXQuantum.2.020318} {\bibfield  {journal} {\bibinfo
   {journal} {PRX Quantum}\ }\textbf {\bibinfo {volume} {2}},\ \bibinfo {pages}
  {020318} (\bibinfo {year} {2021})}\BibitemShut {NoStop}%
\bibitem [{\citenamefont {Sagnol}\ and\ \citenamefont
  {Stahlberg}(2022)}]{PICOS}%
  \BibitemOpen
  \bibfield  {author} {\bibinfo {author} {\bibfnamefont {G.}~\bibnamefont
  {Sagnol}}\ and\ \bibinfo {author} {\bibfnamefont {M.}~\bibnamefont
  {Stahlberg}},\ }\bibfield  {title} {\bibinfo {title} {{PICOS}: A {Python}
  interface to conic optimization solvers},\ }\href
  {https://doi.org/10.21105/joss.03915} {\bibfield  {journal} {\bibinfo
  {journal} {J. Open Source Softw.}\ }\textbf {\bibinfo {volume} {7}},\
  \bibinfo {pages} {3915} (\bibinfo {year} {2022})}\BibitemShut {NoStop}%
\bibitem [{\citenamefont {Stinespring}(1955)}]{Stinespring1955}%
  \BibitemOpen
  \bibfield  {author} {\bibinfo {author} {\bibfnamefont {W.~F.}\ \bibnamefont
  {Stinespring}},\ }\bibfield  {title} {\bibinfo {title} {Positive functions on
  ${C}^*$-algebras},\ }\href
  {https://doi.org/10.1090/s0002-9939-1955-0069403-4} {\bibfield  {journal}
  {\bibinfo  {journal} {Proc. Amer. Math. Soc.}\ }\textbf {\bibinfo {volume}
  {6}},\ \bibinfo {pages} {211} (\bibinfo {year} {1955})}\BibitemShut {NoStop}%
\bibitem [{\citenamefont {Kraus}\ \emph {et~al.}(1983)\citenamefont {Kraus},
  \citenamefont {B\"{o}hm}, \citenamefont {Dollard},\ and\ \citenamefont
  {Wootters}}]{Kraus1983}%
  \BibitemOpen
  \bibinfo {editor} {\bibfnamefont {K.}~\bibnamefont {Kraus}}, \bibinfo
  {editor} {\bibfnamefont {A.}~\bibnamefont {B\"{o}hm}}, \bibinfo {editor}
  {\bibfnamefont {J.~D.}\ \bibnamefont {Dollard}},\ and\ \bibinfo {editor}
  {\bibfnamefont {W.~H.}\ \bibnamefont {Wootters}},\ eds.,\ \href
  {https://doi.org/10.1007/3-540-12732-1} {\emph {\bibinfo {title} {States,
  Effects, and Operations Fundamental Notions of Quantum Theory}}}\ (\bibinfo
  {publisher} {Springer Berlin Heidelberg},\ \bibinfo {year}
  {1983})\BibitemShut {NoStop}%
\bibitem [{\citenamefont {Wilde}(2017)}]{Wilde2017}%
  \BibitemOpen
  \bibfield  {author} {\bibinfo {author} {\bibfnamefont {M.~M.}\ \bibnamefont
  {Wilde}},\ }\bibinfo {title} {Preface to the second edition},\ in\ \href
  {https://doi.org/10.1017/9781316809976.001} {\emph {\bibinfo {booktitle}
  {Quantum Information Theory}}}\ (\bibinfo  {publisher} {Cambridge University
  Press},\ \bibinfo {year} {2017})\ pp.\ \bibinfo {pages} {xi--xii},\ \bibinfo
  {edition} {2nd}\ ed.\BibitemShut {Stop}%
\bibitem [{\citenamefont {Gallego}\ and\ \citenamefont
  {Aolita}(2015)}]{Rodrigo2015}%
  \BibitemOpen
  \bibfield  {author} {\bibinfo {author} {\bibfnamefont {R.}~\bibnamefont
  {Gallego}}\ and\ \bibinfo {author} {\bibfnamefont {L.}~\bibnamefont
  {Aolita}},\ }\bibfield  {title} {\bibinfo {title} {Resource theory of
  steering},\ }\href {https://doi.org/10.1103/PhysRevX.5.041008} {\bibfield
  {journal} {\bibinfo  {journal} {Phys. Rev. X}\ }\textbf {\bibinfo {volume}
  {5}},\ \bibinfo {pages} {041008} (\bibinfo {year} {2015})}\BibitemShut
  {NoStop}%
\bibitem [{\citenamefont {Nery}\ \emph {et~al.}(2020)\citenamefont {Nery},
  \citenamefont {Taddei}, \citenamefont {Sahium}, \citenamefont {Walborn},
  \citenamefont {Aolita},\ and\ \citenamefont {Aguilar}}]{Nery2020}%
  \BibitemOpen
  \bibfield  {author} {\bibinfo {author} {\bibfnamefont {R.~V.}\ \bibnamefont
  {Nery}}, \bibinfo {author} {\bibfnamefont {M.~M.}\ \bibnamefont {Taddei}},
  \bibinfo {author} {\bibfnamefont {P.}~\bibnamefont {Sahium}}, \bibinfo
  {author} {\bibfnamefont {S.~P.}\ \bibnamefont {Walborn}}, \bibinfo {author}
  {\bibfnamefont {L.}~\bibnamefont {Aolita}},\ and\ \bibinfo {author}
  {\bibfnamefont {G.~H.}\ \bibnamefont {Aguilar}},\ }\bibfield  {title}
  {\bibinfo {title} {Distillation of quantum steering},\ }\href
  {https://doi.org/10.1103/PhysRevLett.124.120402} {\bibfield  {journal}
  {\bibinfo  {journal} {Phys. Rev. Lett.}\ }\textbf {\bibinfo {volume} {124}},\
  \bibinfo {pages} {120402} (\bibinfo {year} {2020})}\BibitemShut {NoStop}%
\bibitem [{\citenamefont {Gupta}\ \emph {et~al.}(2021)\citenamefont {Gupta},
  \citenamefont {Das},\ and\ \citenamefont {Majumdar}}]{Gupta2021}%
  \BibitemOpen
  \bibfield  {author} {\bibinfo {author} {\bibfnamefont {S.}~\bibnamefont
  {Gupta}}, \bibinfo {author} {\bibfnamefont {D.}~\bibnamefont {Das}},\ and\
  \bibinfo {author} {\bibfnamefont {A.~S.}\ \bibnamefont {Majumdar}},\
  }\bibfield  {title} {\bibinfo {title} {Distillation of genuine tripartite
  einstein-podolsky-rosen steering},\ }\href
  {https://doi.org/10.1103/PhysRevA.104.022409} {\bibfield  {journal} {\bibinfo
   {journal} {Phys. Rev. A}\ }\textbf {\bibinfo {volume} {104}},\ \bibinfo
  {pages} {022409} (\bibinfo {year} {2021})}\BibitemShut {NoStop}%
\bibitem [{\citenamefont {Ku}\ \emph {et~al.}(2022{\natexlab{b}})\citenamefont
  {Ku}, \citenamefont {Hsieh}, \citenamefont {Chen}, \citenamefont {Chen},\
  and\ \citenamefont {Budroni}}]{Ku2022}%
  \BibitemOpen
  \bibfield  {author} {\bibinfo {author} {\bibfnamefont {H.-Y.}\ \bibnamefont
  {Ku}}, \bibinfo {author} {\bibfnamefont {C.-Y.}\ \bibnamefont {Hsieh}},
  \bibinfo {author} {\bibfnamefont {S.-L.}\ \bibnamefont {Chen}}, \bibinfo
  {author} {\bibfnamefont {Y.-N.}\ \bibnamefont {Chen}},\ and\ \bibinfo
  {author} {\bibfnamefont {C.}~\bibnamefont {Budroni}},\ }\bibfield  {title}
  {\bibinfo {title} {Complete classification of steerability under local
  filters and its relation with measurement incompatibility},\ }\href
  {https://doi.org/10.1038/s41467-022-32466-y} {\bibfield  {journal} {\bibinfo
  {journal} {Nat. Commun.}\ }\textbf {\bibinfo {volume} {13}},\ \bibinfo
  {pages} {4973} (\bibinfo {year} {2022}{\natexlab{b}})}\BibitemShut {NoStop}%
\bibitem [{\citenamefont {Ku}\ \emph {et~al.}(2020)\citenamefont {Ku},
  \citenamefont {Lambert}, \citenamefont {Chan}, \citenamefont {Emary},
  \citenamefont {Chen},\ and\ \citenamefont {Nori}}]{Ku2020}%
  \BibitemOpen
  \bibfield  {author} {\bibinfo {author} {\bibfnamefont {H.-Y.}\ \bibnamefont
  {Ku}}, \bibinfo {author} {\bibfnamefont {N.}~\bibnamefont {Lambert}},
  \bibinfo {author} {\bibfnamefont {F.-J.}\ \bibnamefont {Chan}}, \bibinfo
  {author} {\bibfnamefont {C.}~\bibnamefont {Emary}}, \bibinfo {author}
  {\bibfnamefont {Y.-N.}\ \bibnamefont {Chen}},\ and\ \bibinfo {author}
  {\bibfnamefont {F.}~\bibnamefont {Nori}},\ }\bibfield  {title} {\bibinfo
  {title} {{Experimental test of non-macrorealistic cat states in the cloud}},\
  }\href {https://doi.org/10.1038/s41534-020-00321-x} {\bibfield  {journal}
  {\bibinfo  {journal} {npj Quantum Inf.}\ }\textbf {\bibinfo {volume} {6}},\
  \bibinfo {pages} {98} (\bibinfo {year} {2020})}\BibitemShut {NoStop}%
\bibitem [{\citenamefont {Huang}\ \emph {et~al.}(2021)\citenamefont {Huang},
  \citenamefont {Lin}, \citenamefont {Ku},\ and\ \citenamefont
  {Chen}}]{Huang2021}%
  \BibitemOpen
  \bibfield  {author} {\bibinfo {author} {\bibfnamefont {Y.-T.}\ \bibnamefont
  {Huang}}, \bibinfo {author} {\bibfnamefont {J.-D.}\ \bibnamefont {Lin}},
  \bibinfo {author} {\bibfnamefont {H.-Y.}\ \bibnamefont {Ku}},\ and\ \bibinfo
  {author} {\bibfnamefont {Y.-N.}\ \bibnamefont {Chen}},\ }\bibfield  {title}
  {\bibinfo {title} {Benchmarking quantum state transfer on quantum devices},\
  }\href {https://doi.org/10.1103/PhysRevResearch.3.023038} {\bibfield
  {journal} {\bibinfo  {journal} {Phys. Rev. Research}\ }\textbf {\bibinfo
  {volume} {3}},\ \bibinfo {pages} {023038} (\bibinfo {year}
  {2021})}\BibitemShut {NoStop}%
\bibitem [{\citenamefont {Magesan}\ \emph {et~al.}(2011)\citenamefont
  {Magesan}, \citenamefont {Gambetta},\ and\ \citenamefont
  {Emerson}}]{Magesan2010}%
  \BibitemOpen
  \bibfield  {author} {\bibinfo {author} {\bibfnamefont {E.}~\bibnamefont
  {Magesan}}, \bibinfo {author} {\bibfnamefont {J.~M.}\ \bibnamefont
  {Gambetta}},\ and\ \bibinfo {author} {\bibfnamefont {J.}~\bibnamefont
  {Emerson}},\ }\bibfield  {title} {\bibinfo {title} {Scalable and robust
  randomized benchmarking of quantum processes},\ }\href
  {https://doi.org/10.1103/PhysRevLett.106.180504} {\bibfield  {journal}
  {\bibinfo  {journal} {Phys. Rev. Lett.}\ }\textbf {\bibinfo {volume} {106}},\
  \bibinfo {pages} {180504} (\bibinfo {year} {2011})}\BibitemShut {NoStop}%
\bibitem [{\citenamefont {Magesan}\ \emph {et~al.}(2012)\citenamefont
  {Magesan}, \citenamefont {Gambetta},\ and\ \citenamefont
  {Emerson}}]{Magesan2012}%
  \BibitemOpen
  \bibfield  {author} {\bibinfo {author} {\bibfnamefont {E.}~\bibnamefont
  {Magesan}}, \bibinfo {author} {\bibfnamefont {J.~M.}\ \bibnamefont
  {Gambetta}},\ and\ \bibinfo {author} {\bibfnamefont {J.}~\bibnamefont
  {Emerson}},\ }\bibfield  {title} {\bibinfo {title} {Characterizing quantum
  gates via randomized benchmarking},\ }\href
  {https://doi.org/10.1103/PhysRevA.85.042311} {\bibfield  {journal} {\bibinfo
  {journal} {Phys. Rev. A}\ }\textbf {\bibinfo {volume} {85}},\ \bibinfo
  {pages} {042311} (\bibinfo {year} {2012})}\BibitemShut {NoStop}%
\bibitem [{\citenamefont {Urbanek}\ \emph {et~al.}(2021)\citenamefont
  {Urbanek}, \citenamefont {Nachman}, \citenamefont {Pascuzzi}, \citenamefont
  {He}, \citenamefont {Bauer},\ and\ \citenamefont {de~Jong}}]{Urbanek2021prl}%
  \BibitemOpen
  \bibfield  {author} {\bibinfo {author} {\bibfnamefont {M.}~\bibnamefont
  {Urbanek}}, \bibinfo {author} {\bibfnamefont {B.}~\bibnamefont {Nachman}},
  \bibinfo {author} {\bibfnamefont {V.~R.}\ \bibnamefont {Pascuzzi}}, \bibinfo
  {author} {\bibfnamefont {A.}~\bibnamefont {He}}, \bibinfo {author}
  {\bibfnamefont {C.~W.}\ \bibnamefont {Bauer}},\ and\ \bibinfo {author}
  {\bibfnamefont {W.~A.}\ \bibnamefont {de~Jong}},\ }\bibfield  {title}
  {\bibinfo {title} {Mitigating depolarizing noise on quantum computers with
  noise-estimation circuits},\ }\href
  {https://doi.org/10.1103/PhysRevLett.127.270502} {\bibfield  {journal}
  {\bibinfo  {journal} {Phys. Rev. Lett.}\ }\textbf {\bibinfo {volume} {127}},\
  \bibinfo {pages} {270502} (\bibinfo {year} {2021})}\BibitemShut {NoStop}%
\bibitem [{\citenamefont {Figueroa-Romero}\ \emph {et~al.}(2021)\citenamefont
  {Figueroa-Romero}, \citenamefont {Modi}, \citenamefont {Harris},
  \citenamefont {Stace},\ and\ \citenamefont {Hsieh}}]{Romero2021prx}%
  \BibitemOpen
  \bibfield  {author} {\bibinfo {author} {\bibfnamefont {P.}~\bibnamefont
  {Figueroa-Romero}}, \bibinfo {author} {\bibfnamefont {K.}~\bibnamefont
  {Modi}}, \bibinfo {author} {\bibfnamefont {R.~J.}\ \bibnamefont {Harris}},
  \bibinfo {author} {\bibfnamefont {T.~M.}\ \bibnamefont {Stace}},\ and\
  \bibinfo {author} {\bibfnamefont {M.-H.}\ \bibnamefont {Hsieh}},\ }\bibfield
  {title} {\bibinfo {title} {Randomized benchmarking for non-markovian noise},\
  }\href {https://doi.org/10.1103/PRXQuantum.2.040351} {\bibfield  {journal}
  {\bibinfo  {journal} {PRX Quantum}\ }\textbf {\bibinfo {volume} {2}},\
  \bibinfo {pages} {040351} (\bibinfo {year} {2021})}\BibitemShut {NoStop}%
\bibitem [{\citenamefont {Ebler}\ \emph {et~al.}(2018)\citenamefont {Ebler},
  \citenamefont {Salek},\ and\ \citenamefont {Chiribella}}]{Ebler2018}%
  \BibitemOpen
  \bibfield  {author} {\bibinfo {author} {\bibfnamefont {D.}~\bibnamefont
  {Ebler}}, \bibinfo {author} {\bibfnamefont {S.}~\bibnamefont {Salek}},\ and\
  \bibinfo {author} {\bibfnamefont {G.}~\bibnamefont {Chiribella}},\ }\bibfield
   {title} {\bibinfo {title} {Enhanced communication with the assistance of
  indefinite causal order},\ }\href
  {https://doi.org/10.1103/PhysRevLett.120.120502} {\bibfield  {journal}
  {\bibinfo  {journal} {Phys. Rev. Lett.}\ }\textbf {\bibinfo {volume} {120}},\
  \bibinfo {pages} {120502} (\bibinfo {year} {2018})}\BibitemShut {NoStop}%
\bibitem [{\citenamefont {Loizeau}\ and\ \citenamefont
  {Grinbaum}(2020)}]{Loizeau2020}%
  \BibitemOpen
  \bibfield  {author} {\bibinfo {author} {\bibfnamefont {N.}~\bibnamefont
  {Loizeau}}\ and\ \bibinfo {author} {\bibfnamefont {A.}~\bibnamefont
  {Grinbaum}},\ }\bibfield  {title} {\bibinfo {title} {Channel capacity
  enhancement with indefinite causal order},\ }\href
  {https://doi.org/10.1103/PhysRevA.101.012340} {\bibfield  {journal} {\bibinfo
   {journal} {Phys. Rev. A}\ }\textbf {\bibinfo {volume} {101}},\ \bibinfo
  {pages} {012340} (\bibinfo {year} {2020})}\BibitemShut {NoStop}%
\bibitem [{\citenamefont {Felce}\ and\ \citenamefont
  {Vedral}(2020)}]{Felce2020}%
  \BibitemOpen
  \bibfield  {author} {\bibinfo {author} {\bibfnamefont {D.}~\bibnamefont
  {Felce}}\ and\ \bibinfo {author} {\bibfnamefont {V.}~\bibnamefont {Vedral}},\
  }\bibfield  {title} {\bibinfo {title} {Quantum refrigeration with indefinite
  causal order},\ }\href {https://doi.org/10.1103/PhysRevLett.125.070603}
  {\bibfield  {journal} {\bibinfo  {journal} {Phys. Rev. Lett.}\ }\textbf
  {\bibinfo {volume} {125}},\ \bibinfo {pages} {070603} (\bibinfo {year}
  {2020})}\BibitemShut {NoStop}%
\bibitem [{\citenamefont {Guha}\ \emph {et~al.}(2020)\citenamefont {Guha},
  \citenamefont {Alimuddin},\ and\ \citenamefont {Parashar}}]{Guha2020}%
  \BibitemOpen
  \bibfield  {author} {\bibinfo {author} {\bibfnamefont {T.}~\bibnamefont
  {Guha}}, \bibinfo {author} {\bibfnamefont {M.}~\bibnamefont {Alimuddin}},\
  and\ \bibinfo {author} {\bibfnamefont {P.}~\bibnamefont {Parashar}},\
  }\bibfield  {title} {\bibinfo {title} {Thermodynamic advancement in the
  causally inseparable occurrence of thermal maps},\ }\href
  {https://doi.org/10.1103/PhysRevA.102.032215} {\bibfield  {journal} {\bibinfo
   {journal} {Phys. Rev. A}\ }\textbf {\bibinfo {volume} {102}},\ \bibinfo
  {pages} {032215} (\bibinfo {year} {2020})}\BibitemShut {NoStop}%
\bibitem [{\citenamefont {Simonov}\ \emph {et~al.}(2022)\citenamefont
  {Simonov}, \citenamefont {Francica}, \citenamefont {Guarnieri},\ and\
  \citenamefont {Paternostro}}]{Simonov2022}%
  \BibitemOpen
  \bibfield  {author} {\bibinfo {author} {\bibfnamefont {K.}~\bibnamefont
  {Simonov}}, \bibinfo {author} {\bibfnamefont {G.}~\bibnamefont {Francica}},
  \bibinfo {author} {\bibfnamefont {G.}~\bibnamefont {Guarnieri}},\ and\
  \bibinfo {author} {\bibfnamefont {M.}~\bibnamefont {Paternostro}},\
  }\bibfield  {title} {\bibinfo {title} {Work extraction from coherently
  activated maps via quantum switch},\ }\href
  {https://doi.org/10.1103/PhysRevA.105.032217} {\bibfield  {journal} {\bibinfo
   {journal} {Phys. Rev. A}\ }\textbf {\bibinfo {volume} {105}},\ \bibinfo
  {pages} {032217} (\bibinfo {year} {2022})}\BibitemShut {NoStop}%
\end{thebibliography}
\end{document}